\documentclass[epj]{svjour}
\usepackage{amsmath,amssymb,epsfig,bm,mathrsfs,slashed,color,graphicx,mathtools,isotope,aas_macros}
\usepackage[numbers,sort&compress]{natbib}
\usepackage{filecontents}
\usepackage{tikz}
\usepackage[utf8]{inputenc}
\newcommand{\bea}{\begin{eqnarray}}
\newcommand{\eea}{\end{eqnarray}}
\newcommand{\be}{\begin{equation}}
\newcommand{\ee}{\end{equation}}

\newcommand{\vectau}{{\bm \tau}}
\newcommand{\vecrho}{{\bm n}}

\newcommand{\Lsym}{L_{\rm sym}}
\newcommand{\Qsat}{Q_{\rm sat}}
\definecolor{red}{rgb}{0.8,0,0}
\definecolor{violet}{rgb}{0.4,0,0.4}
\definecolor{green}{rgb}{0,0.5,0.0}
\definecolor{navy}{rgb}{0.0,0.0,0.6}
\definecolor{orange}{rgb}{0.8,0.2,0.0}

\usepackage[normalem]{ulem}  

\usepackage[colorlinks=true]{hyperref}
\definecolor{reddish}{rgb}{0.7,0.2,0.0}
\definecolor{blueish}{rgb}{0.1,0.1,1}

\hypersetup{
  linkcolor=reddish,          
  citecolor=blueish,           
urlcolor=magenta}            
\journalname{Eur. Phys. J. A}

\begin{document}

\title{Finite-temperature equations of state  of compact stars  with hyperons: three-dimensional tables}

\author{Stefanos Tsiopelas$^1$,  Armen Sedrakian$^{1,2}$, Micaela Oertel$^{3}$
}                     
\institute{
$^1$   Institute of Theoretical Physics, University of Wroc\l{}aw, pl. M. Borna 9, 
   50-204 Wroc\l{}aw, Poland\\
$^2$  Frankfurt Institute for Advanced Studies, D-60438 Frankfurt am Main, Germany\\
$^3$   LUTH, Observatoire de Paris, Universit\'e PSL, CNRS, Universit\'e Paris Cit\'e, 92190 Meudon, France
}
\date{Received: 17 April 2024  / Accepted: 23 May 2024\\
\textcopyright \, 
The Author(s),  under exclusive licence to Società Italiana di Fisica and Springer-Verlag GmbH Germany, part of Springer Nature  2024
\\
Communicated by David Blaschke
}
%
\abstract{ We construct tables of finite temperature equation of state
  (EoS) of hypernuclear matter in the range of densities,
  temperatures, and electron fractions that are needed for numerical
  simulations of supernovas, proto-neutron stars, and binary neutron
  star mergers and cast them in the format of {\sc CompOSE}
  database. The tables are extracted from a model that is based on
  covariant density functional (CDF) theory that includes the full
  $J^P=1/2^+$ baryon octet in a manner that is consistent with the
  current astrophysical and nuclear constraints.  We employ a
  parameterization with three different values of the slope of the
  symmetry energy $\Lsym=30$, 50 and 70 MeV and fixed skewness $\Qsat=
  400$ MeV for above saturation matter.  A model for the EoS of
  inhomogeneous matter is matched at sub-saturation density to the
  high-density hypernuclear EoS. We discuss the generic features of
  the resulting EoS and the composition of matter as a function of
  density, temperature, and electron fraction.  The nuclear characteristics and strangeness fraction of these models are
  compared to the alternatives from the literature.  
  The integral properties of static and rapidly rotating compact stars in the limit
  of zero temperature are discussed and confronted with the
  multimessenger astrophysical constraints.
} 
\authorrunning{S. Tsiopelas, et al. }
\titlerunning{Three-dimensional tables}
\maketitle
%

\section{Introduction}
\label{sec:Introduction}

The equation of state (EoS) for dense, strongly interacting matter serves as the central input in an array of astrophysical simulations involving isolated compact objects and binary systems across various scenarios. The {\sc  CompOSE}  database~\cite{Typel2022, Dexheimer2022} hosts a substantial collection of EoS data. Nevertheless, while numerous models exist to describe the composition of mature cold neutron stars, the range narrows considerably when considering the so-called general-purpose EoS, which encompass varying temperature, density, and electron fraction. Moreover, particularly for EoS models incorporating non-nucleonic degrees of freedom in dense and hot matter, like hyperons, recent years have imposed stringent astrophysical constraints, rendering several existing models incompatible with observations.  It is expected that more constraints will emerge in the future in light of (multimessenger) observations of binary neutron star (BNS) mergers, isolated X-ray-emitting neutron stars in our proximity, and radio pulsars. In this context, the necessity of the inclusion of new EoS into this and other databases 
allows us to better cover and quantify the large uncertainties in the description of dense and hot strongly interacting matter. This paper describes the generation of general-purpose EoS tables in the temperature, density, and electron fraction space based on covariant density functional (CDF) models which include hyperonic degrees of freedom, for reviews see~\cite{Oertel2017,FiorellaBurgio:2018dga,Sedrakian2023}. 
Specifically, CDFs for finite temperature hypernuclear matter of the kind that we will employ here have been constructed and applied to a range of astrophysical scenarios in the past~\cite{Colucci2013,vanDalen2014,Marques2017,Fortin2018,Raduta2020,Sedrakian2021,Sedrakian2022,Kochankovski2022}. 

Our general aim here is to cover the complete parameter space of temperature, density, and electron fraction,  that is required by the simulations of the astrophysical scenarios mentioned above. We will employ recent new parametrizations of the CDF with density-dependent (DD) couplings~\cite{Li2023} which allow for variations in the slope of the symmetry energy $\Lsym$ and the skewness $\Qsat$. Here, we discuss the EoS with contributions from baryons, photons, and electrons/positrons to the EoS, the two latter components being treated as an ideal gas. The case of neutrino-trapped matter in the high-density range $n_B/n_{\rm sat} \ge 0.5$, where $n_{\rm sat}$  is the saturation density has been discussed in Refs.~\cite{Sedrakian2021,Sedrakian2022} where a different (DDME2) parameterization was 
used~\cite{Lalazissis2005}. We will compare the results from this and other alternative parametrizations to the present results later on.

 We further match the EoS of high-density (homogeneous) hypernuclear matter to that of low-density (inhomogeneous) nuclear matter. The low-density EoS corresponds to the one developed in Ref.~\cite{Hempel2010} and is based on an improved nuclear statistical equilibrium among nucleons and nuclear clusters. Note that the hypernuclear matter is described within the class of CDF models which are consistent with the current astrophysical constraints on cold neutron star radii, tidal deformabilities, and maximum masses - models that emerged mostly after the 2010 detection of the 2-solar mass compact star, for a discussion see~\cite{Sedrakian2023}.  The model we employ includes the full baryon octet, i.e., strangeness 1 and 2 hyperons. It allows for  $\sigma^*$ and $\phi$ hidden strangeness mesons which mediate the hyperon-hyperon interaction.

Numerical simulations reveal that BNS mergers generate hot and dense strongly interacting matter during the post-merger stage~\cite{Rosswog2015, Baiotti2017, Baiotti2019,Blacker2024}. The outcome of such a merger is contingent upon the combined masses of the merging objects and may lead to the formation of a black hole or a stable neutron star. In any scenario, a transient hot object is formed, and thus, the spectrum of gravitational waves emitted during this phase (potentially observable with advanced gravitational wave instruments, such as the Einstein Telescope~\cite{Branchesi:2023mws}) carries distinctive characteristics reflecting the EoS of hot and dense matter. This EoS also governs the stability of the resulting object, influencing the course of its transient evolution~\cite{Khadkikar2021}, as well as the efficiency of dissipative processes~\cite{Alford2019a, Alford2019b, Alford2020, Alford2021a, Alford2021b, Alford2022, Most_2024, Celora2022}, which should be incorporated into the commonly used ideal hydrodynamics simulations~\cite{Most2022}.

The emergence of hot and neutrino-rich compact objects is also anticipated through numerical simulations of core-collapse supernovas (CCSN). As the supernova progenitor contracts, a hot proto-neutron star is formed along with expanding ejecta~\cite{Prakash1997, Pons1999, Janka2007, Foglizzo2015, Connor2018, Burrows2020, Pascal2022, Mezzacappa2023}. The transient formation of dense, high-temperature matter occurs also when the progenitor possesses significant mass and the material collapses ultimately into a black hole~\cite{Sumiyoshi2007, Fischer2009, Connor2011, Schneider2020}.

The local characteristics of matter during the hot phase in the aforementioned astrophysical scenarios are primarily defined by factors such as density, temperature (or entropy), and the lepton fractions for electrons and muons. The EoS in this phase depends on multiple parameters, which can be compared to the more straightforward one-parameter EoS describing cold and $\beta$-equilibrated matter. Given the multitude of nuclear and astrophysical constraints imposed on the cold EoS we will briefly touch upon the predictions of underlying EoS for global properties of cold nucleonic and hypernuclear compact stars; such input is also required by the  {\sc CompOSE}  repository. Muons are excluded from our tables to maintain a three-dimensional representation of parameter space, focusing on temperature, density, and electron fraction.

This work is organized as follows. In Section~\ref{sec:EoS_high_density} we review the high-density EoS of hypernuclear matter at finite temperatures, where Subsection \ref{subsec:formalism} discusses the formalism, Subsec.~\ref{subsec:Couplings} - the choice of the coupling constants,   Subsec.~\ref{subsec:Adapting_to_Astro} the conditions relevant for BNS mergers and CCSN and Subsec.~\ref{subsec:Matching} the procedure of 
matching the low (inhomogeneous) and high-density (homogeneous) EoS.  
We present the numerical results on the EoS and composition in Section~\ref{sec:Num_EoS_Composition}. The mass-radius (hereafter  $M$-$R$) relation of static cold compact stars and the astrophysical constraints are discussed in Section~\ref{sec:MR}. Section~\ref{sec:Comparison} is dedicated to the discussion of how our model compares to the alternative models. In Section~\ref{sec:Conclusions} we provide a summary of our main results. We use the natural (Gaussian) units with $\hbar= c=k_B=1$, and the metric signature 
  $g^{\mu\nu}={\rm diag}(1,-1,-1,-1)$.

\section{Finite temperature equation of state of hypernuclear matter}
\label{sec:EoS_high_density}

The purpose of this section is to collect all the ingredients that are required for the construction of the three-dimensional tables. We review the physical model, the selection of the values for the couplings in the Lagrangian of the model, the thermodynamical conditions relevant for the cases of BNS and CCSN, and, finally, the matching to the low-density (subnuclear) EoS to the high-density one. 

\subsection{Formalism}
\label{subsec:formalism}

The CDF model on which the finite-temperature EoSs is constructed is based on the Lagrangian 
\begin{equation}
  \label{eq:Lagrangian}
  \mathscr{L} = \mathscr{L}_b  + \mathscr{L}_m +  \mathscr{L}_\lambda
  +  \mathscr{L}_{\rm em},
\end{equation}
where the $J_B^P = \frac{1}{2}^+$  baryon Lagrangian is given by 
\begin{eqnarray}
  \label{eq:Lagrangian_b} 
  \mathscr{L}_b \, &=&\,  \sum_b\bar\psi_b\bigg[\gamma^\mu 
  \left(i\partial_\mu-g_{\omega b}\omega_\mu
  -g_{\phi b}\phi_\mu - \frac{1}{2} g_{n
  b}\vectau\cdot\vecrho_\mu\right) \nonumber\\
  &-& (m_b - g_{\sigma b}\sigma 
  - g_{\sigma^* b}\sigma^*)\bigg]\psi_b ,
\end{eqnarray}
 with the index $b$ summing over the $J_B^P = \frac{1}{2}^+$ baryon octet, 
 which includes neutrons, protons,  and $\Lambda,$ $\Xi^{0,-}$, $\Sigma^{0,\pm}$ hyperons. 
 The interaction is modeled via the non-strange sector mesons $\sigma,\omega_\mu,$ and $\vecrho_\mu$ which couple to all the members of the octet and hidden strangeness mesons 
$\sigma^*,\phi_\mu$, which couple
only to hyperons.  Here  $\psi_b$ are the Dirac
fields of the baryon octet with masses $m_b$, $g_{mb}$ are the 
meson-baryon couplings where $m$ index runs over the different meson channels.
Note that the meson states here do not correspond to the ones in vacuum, but effectively describe the interaction of baryons taking into account multiple scattering and in-medium effects.  It is assumed that the masses are close to those in vacuum, except the $\sigma$ meson, which has a resonance structure and effectively represents two-pion state. Its mass is often used as a free parameter of a model.
  Their couplings to baryons, however, 
 differ from the tree-level couplings in vacuum physics, i.e., are treated as fit parameters that account for the above-mentioned multiple scattering and in-medium effects.

The mesonic part of the Lagrangian is given by
\begin{eqnarray}
\mathscr{L}_m &=& \frac{1}{2}
\partial^\mu\sigma\partial_\mu\sigma-\frac{m_\sigma^2}{2} \sigma^2 -
                  \frac{1}{4}\omega^{\mu\nu}\omega_{\mu\nu}
                  + \frac{m_\omega^2}{2}
                  \omega^\mu\omega_\mu \nonumber\\
  &-& \frac{1}{4}\vecrho^{\mu\nu}\cdot \vecrho_{\mu\nu}
               + \frac{m_n^2}{2} \vecrho^\mu\cdot\vecrho_\mu 
               +\frac{1}{2}
\partial^\mu\sigma^*\partial_\mu\sigma^*-\frac{m_\sigma^{*2}}{2}
  \sigma^{*2}\nonumber\\
  &-&
\frac{1}{4}\phi^{\mu\nu}\phi_{\mu\nu} + \frac{m_\phi^2}{2}
               \phi^\mu\phi_\mu,
\end{eqnarray}
where $m_{\sigma}$, $m_{\sigma^*}$, $m_{\omega}$, $m_{\phi}$ and  $m_{n}$ denote the meson masses. 
The field-strength tensors for vector
fields are given by 
\begin{eqnarray}             
\omega_{\mu \nu}  &=& \partial_{\mu}\omega_{\nu} - \partial_{\mu}\omega_{\nu} ,\\
\phi_{\mu \nu} & =& \partial_{\mu}\phi_{\nu} - \partial_{\mu}\phi_{\nu} ,\\
\boldsymbol{n}_{\mu \nu}  &=& \partial_{\nu}
\boldsymbol{n}_{\mu} - \partial_{\mu}\boldsymbol{n}_{\nu}.
\end{eqnarray}
As mentioned above, charged leptons are treated as an ideal gas, thus the Lagrangian is given by  the free Dirac Lagrangian
\begin{eqnarray}
\label{eq:Lagrangian_leptons}
\mathscr{L}_\lambda\,=\, \sum_{\lambda}\bar\psi_\lambda(i\gamma^\mu\partial_\mu -
      m_\lambda)\psi_\lambda,
\end{eqnarray}
where $\psi_\lambda$ are leptonic fields and $m_\lambda$ are their masses. 

The general-purpose tables of EoS commonly include only electrons and positrons and neglect the contributions from other leptons. Charged $\tau$-leptons are too massive to appear in stellar matter. Including charged muons treated as an ideal gas in an EoS table is a priori straightforward since the main difficulty resides in determining the EoS for the strongly interacting baryons. Since in addition only a few simulations up to now evolve a muon fraction, see e.g.~\cite{Bollig:2017lki}, we refrain here from adding a fourth dimension to our already rather memory-consuming tables. A similar reasoning applies to the contributions of neutrinos which in the trapped regime contribute to energy density and pressure. These contributions can be cast into analytical expressions which can easily be added if needed. Note that in the case when strong electromagnetic fields are present \eqref{eq:Lagrangian} should include 
the electromagnetic contribution $\mathscr{L}_{\rm em}$ which we neglect here. Examples, where
such term is important are the cores of magnetars that may  support magnetic fields of the order of $10^{18}$~G~\cite{Chatterjee2015,Thapa2020,Peterson2023}as well as magnetar crusts where the physics of finite nuclei may be affected by magnetic fields~\cite{Stein2016,Scurto2023,Parmar2023,Kondratyev2020}. Electromagnetic part of the Largangian contributes also to the Coulomb corrections in inhomogeneous matter, see e.g.~Ref.~\cite{Hempel2010},
but these are not relevant for present discussion.
For the Lagrangian given by Eq.~\eqref{eq:Lagrangian} the evaluation of the pressure and energy density of the constituents can be added with $\lambda=e$
\begin{eqnarray}
  \label{eq:P}
  P  &=&  P_b+P_m+P_{e} + P_{r},\\
    \label{eq:E}
          {\cal E} &=& {\cal E}_b+ {\cal E}_m+{\cal E}_{e},
\end{eqnarray}
where the contributions due to mesons and  $J_B^P =
\frac{1}{2}^+$-baryons are given by
\begin{eqnarray}
    \label{eq:P_m}
P_m &=& - \frac{m_\sigma^2}{2} \sigma^2 -\frac{m_\sigma^{*2}}{2} \sigma^{*2}
          + \frac{m_\omega^2}{2} \omega_0^2 +  \frac{m_\phi^2 }{2} \phi_0^2
        + \frac{m_n^2 }{2} n_{03}^2,\nonumber\\ \\
    \label{eq:E_m}
{\cal E}_m &=& \frac{m_\sigma^2}{2} \sigma^2 +\frac{m_\sigma^{*2}}{2} \sigma^{*2} +
             \frac{m_\omega^2 }{2} \omega_0^2 
                          +  \frac{m_\phi^2 }{2} \phi_0^2
                          + \frac{m_n^2 }{2} n_{03}^2,\\
    \label{eq:P_b}
  P_b&=& \sum_{b} \frac{g_{b}}{6\pi^2}\int_0^{\infty}\!
  \frac{dk\, k^4 }{E_k^{b}}
     \left[f(E_k^{b}-\mu_{b}^*)+f(E_k^{b}+\mu_{b}^*)\right],\nonumber\\\\
    \label{eq:E_b}
    {\cal E}_b &=& \sum_{b} \frac{g_{b}}{2\pi^2}  
\int_0^{\infty}\! dk \, k^2 E_k^{b} \left[f(E_k^{b}-\mu_{b}^*)
                   +f(E_k^{b}+\mu_{b}^*)\right] ,\nonumber\\
\end{eqnarray}
where $f(E) = [1+\exp(E/T)]^{-1}$ is the Fermi distribution function at temperature $T$, the
single-particle energies of baryons  are given by $E_k^{b} = \sqrt{k^2+m^{*2}_{b}}$, 
$g_b=2J_b+1 = 2 $ is the spin ($J_b = 1/2$) degeneracy factor of the baryon octet, and the effective masses are defined as 
\begin{equation}
  m_{b}^* = m_b - g_{\sigma b}\sigma - g_{\sigma^*b}\sigma^*.
\end{equation}
The effective chemical potentials are given by
\begin{eqnarray}
\label{eq:mu_eff}
&& \mu_{b}^* = \mu_{b}- g_{\omega b}\omega_{0} - g_{\phi b}\phi_{0} 
   - g_{n b}  n_{03} I_{3b} - \Sigma^{r}, 
\end{eqnarray}
where the so-called rearrangement term is given by 
\begin{eqnarray}
  \Sigma^r&=&\sum_{b,d} \left(
  \frac{\partial g_{\omega b}}{\partial n_b} \omega_0 n_b+
 \frac{\partial g_{n b}}{\partial n_b}  I_{3b} n_{03} n_b+
              \frac{\partial g_{\phi b}}{\partial n_b}  \phi_0 n_b
              \right.\nonumber\\
  &-&\left.
  \frac{ \partial g_{\sigma b}}{\partial n_b} \sigma n_b^s
  -  \frac{ \partial g_{\sigma^* b}}{\partial n_b} \sigma^* n_b^s
  \right)
\end{eqnarray}
and the pressure contribution from this term is 
$P_r = n_B\Sigma_{r}$. Note that we implicitly used the fact that the baryon-meson couplings depend on the vector density, therefore the rearrangement term appears in Eq.~\eqref{eq:mu_eff} for the chemical potentials. If, however, scalar-density dependence is used, as in the MPE model of Ref.~~\cite{Typel2018}, then the formalism needs some modifications \cite{Li2023}.
The mesonic fields in Eqs.~\eqref{eq:P_m}, \eqref{eq:E_m} 
and the following equations correspond to the mean-field values.
Their expectation values in the infinite system
approximations are standard and we will not write them down here. 
The system of equations for baryons is closed by the expressions for
the scalar and baryon (vector) number densities:
\bea\label{eq:density_b}
n_{b} &=& \frac{g_b}{2\pi^2}\!\!\int_0^\infty\!   k^2dk
\left[f(E^b_k-\mu_b^*)-f(E^b_k+\mu_b^*)\right],\\
\label{eq:density_s}
n_{b}^s &=& \frac{g_b}{2\pi^2}\!\!\int_0^\infty\! \frac{k^2dk\, m^*_b}{E_k^b} 
\left[f(E^b_k-\mu_b^*)+f(E^b_k+\mu_b^*)\right].\quad
\eea 
The electronic contribution is given by
\begin{eqnarray}
  P_{e} &=&
  \frac{g_{e}}{6\pi^2}\int_0^{\infty}\!
   \frac{dk\, k^4 }{E_k^e}
   \left[f(E_k^{e}-\mu_e)+f(E_k^{e}+\mu_e)\right],\nonumber\\
  \\
  {\cal E}_{e} &=&                     \frac{g_{e}}{2\pi^2}  \int_0^{\infty}\!\!
  dk\, k^2E_k^e\left[f(E_k^e-\mu_e)
+f(E_k^e+\mu_e)\right], \nonumber\\
\end{eqnarray}
where the degeneracy factor for electrons $g_e=2$.  Electronic energies are given just by their kinetic energy 
$E_k^{e}=\sqrt{k^2+m_e^2}$, where $m_e$ is given by
the free mass of an electron.

\subsection{Coupling constants}
\label{subsec:Couplings}

In total six tables were constructed, three of which are purely nucleonic 
and are based on the parameterization of Ref.~\cite{Li2023}. 
The other three EoS contain in addition hyperons. 
The three nucleonic EoS differ by the value of the slope of the symmetry energy $\Lsym=30$, 50 and 70 MeV and have fixed skewness $\Qsat$; we will abbreviate them as  DDLS(30), DDLS(50), and DDLS(70), respectively. 
 The value $Q_{\rm sat}=400$~MeV was chosen to ensure that the maximum masses of hypernuclear stars surpass the lower bound of two solar masses. The considered range for $30\le L_{\rm sym}\le 70$~MeV accommodates existing uncertainties associated with neutron skin measurements and neutron star radii~\cite{Lattimer2023}. Consequently, our ensemble of EoS facilitates investigations of the impact of varying the symmetry energy slope on dynamic phenomena such as supernova explosions and BNS mergers.

 Let us now recall that the energy density of  nuclear matter  in the vicinity of saturation density and isospin-symmetrical limit can be expressed via a double-expansion in the Taylor series:
\begin{eqnarray}
\label{eq:Taylor_expansion}
  E(\chi, \delta) &\simeq &
  E_{\text{sat}} + \frac{1}{2!}K_{\text{sat}}\chi^2
  + \frac{1}{3!}Q_{\text{sat}}\chi^3 \nonumber\\
  &+& \left(J_{\text{sym}} + L_{\text{sym}}\chi +\frac{1}{2!}K_{\text{sym}}\chi^2 + \frac{1}{3!\
}Q_{\text{sym}}\chi^3\right) \delta^2\nonumber\\
\end{eqnarray}
where $\chi=(n-n_{\text{sat}})/3n_{\text{sat}}$ and $\delta = (n_{n}-n_{p})/n$, with $n_n$ and $n_p$ being the neutron and proton densities.

The definition of the coefficients of the expansion are standard and are given, e.g., in Ref.~\cite{Sedrakian2023}. For the density dependence of the nucleon-meson couplings, we assume a standard form~\cite{TypelWolter1999}
\begin{equation}
 g_{iN}(n_B) = g_{iN}(n_{0})h_i(x),
\end{equation}
where  $x = n_B/n_{0}$ and $n_0$ is a reference density specified in the parametrization. 
\begin{eqnarray} \label{eq:h_functions}
  h_i(x) &=&a_i\frac{1+b_i(x+d_i)^2}{1+c_i(x+d_i)^2},~i=\sigma,\omega,\\
  \label{eq:h_function_rho}
h_n(x) &=& e^{-a_n(x-1)}.
\end{eqnarray}
The explicit values of couplings and parameters determining the density-dependence are given in Ref.~\cite{Li2023}.  
To complete the discussion, let us note that the saturation density $n_{\text {sat }}=0.152$ fm$^{-3}$,  the binding energy per particle in symmetrical nuclear matter at saturation density $E_B=-16.14$ MeV, the compressibility $K=251$ MeV and effective nucleon mass $m^*_{n,p}/m_N=0.57$~MeV
(where $m_N=939$~MeV is the bare mass)
for the DDLS models are by construction the same as for the DDME2 model. The symmetry energy $E_{\rm sym}$, however, changes with the value of the $\Lsym$, specifically, $E_{\rm sym}=30.1,\, 32.2, \, 34.0$ MeV for 
$\Lsym=30,\,50\,$ and 70~MeV, respectively.

The density dependence of the couplings for hyperons is the same as those for nucleonic ones, but their strengths at the reference density $n_0$ are different. The ratios of 
the hyperonic to nucleonic ones are given in Table~\ref{tab:1}. The depths  of hyperonic potentials in the symmetric nuclear matter are  
$U_\Lambda (n_{\rm sat})=  -30$~MeV $U_\Xi (n_{\rm sat}) =-14$~MeV
and $U_\Sigma (n_{\rm sat})= +30$~MeV.
\begin{table}[t]
\centering
\caption{The ratios of the couplings of hyperons to mesons to those of nucleons at saturation density, i.e.,  $n_0=n_{\rm sat}$.}
\begin{tabular}{cccccc}
    \hline
 $b\backslash R$  & $R_{\omega b}$  & $R_{\phi b}$ & $R_{n b}$
 & $R_{\sigma b}$  &$R_{\sigma^* b}$\\
  \hline
$\Lambda$ &   2/3 &  $-\sqrt{2}/3$    & 0 & 0.6106 & 0.4777 \\
$\Sigma $ &   2/3 &  $-\sqrt{2}/3$    & 2 & 0.4426 & 0.4777 \\
  $\Xi$     &   1/3 &  $-2\sqrt{2}/3$   & 1 & 0.3024 & 0.9554\\
\hline
\end{tabular}
\label{tab:1}
\end{table}

\subsection{Adapting CDF to conditions in CCSN and BNS merger remnants}
\label{subsec:Adapting_to_Astro}

For zero-temperature computations, we will assume weak-equi\-li\-brium among the members of the baryon octet and electrons. It is assumed that neutrinos freely escape the star and, therefore, do not form a statistical ensemble. This implies the following relations among the chemical potentials 
\begin{eqnarray}
\label{eq:c1}
  &&\mu_{\Lambda}=\mu_{\Sigma^0}=\mu_{\Xi^0}=\mu_n=\mu_B,\\
\label{eq:c2}
 &&  \mu_{\Sigma^-}=\mu_{\Xi^-}=\mu_B-\mu_Q,\\
\label{eq:c3}
  &&\mu_{\Sigma^+}=\mu_B+\mu_Q,
\end{eqnarray}
where $\mu_B$ and $\mu_Q=\mu_p-\mu_n=-\mu_e$ are the baryon and charge
chemical potentials.  An additional constraint is imposed by the charge neutrality condition
    \begin{eqnarray}
    &&  n_p+n_{\Sigma^+}-n_{\Sigma^-}+n_{\Xi^-}-n_{e^-}+n_{e^+}=0.
\end{eqnarray}
where $e^{\pm}$ refers to electrons and positrons, respectively. We will work below with a fixed electron fraction which is given by $Y_{e}=(n_{e^-}-n_{e^+})/n_B$.
\begin{figure*}[!tbp] 
\begin{center}
\includegraphics[width=\linewidth,keepaspectratio]{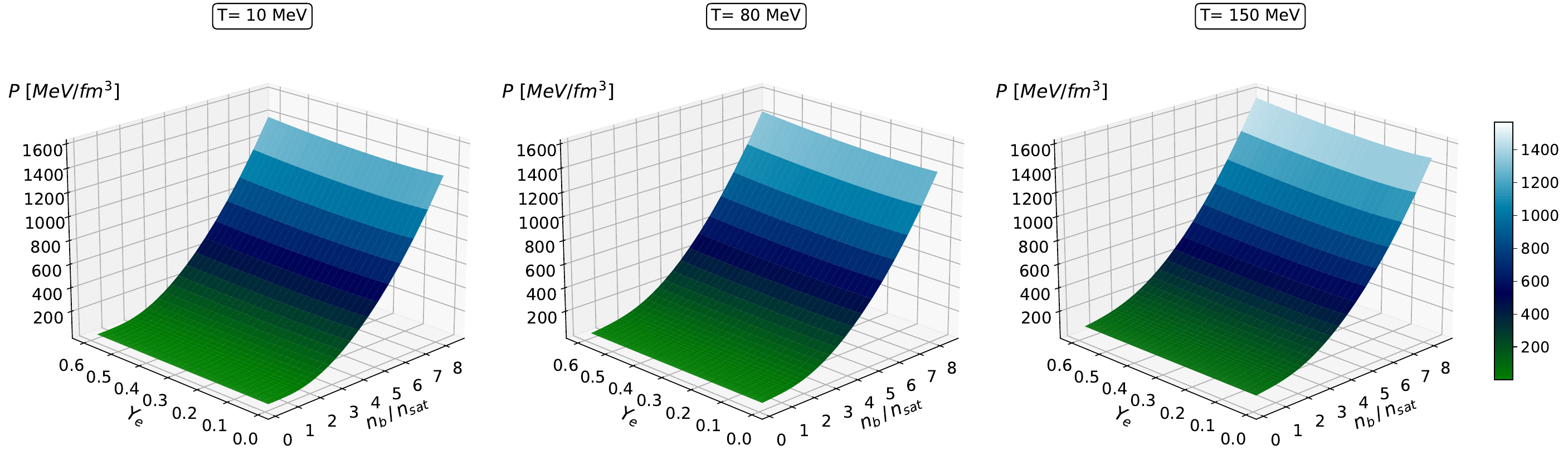}
\includegraphics[width=\linewidth,keepaspectratio]
{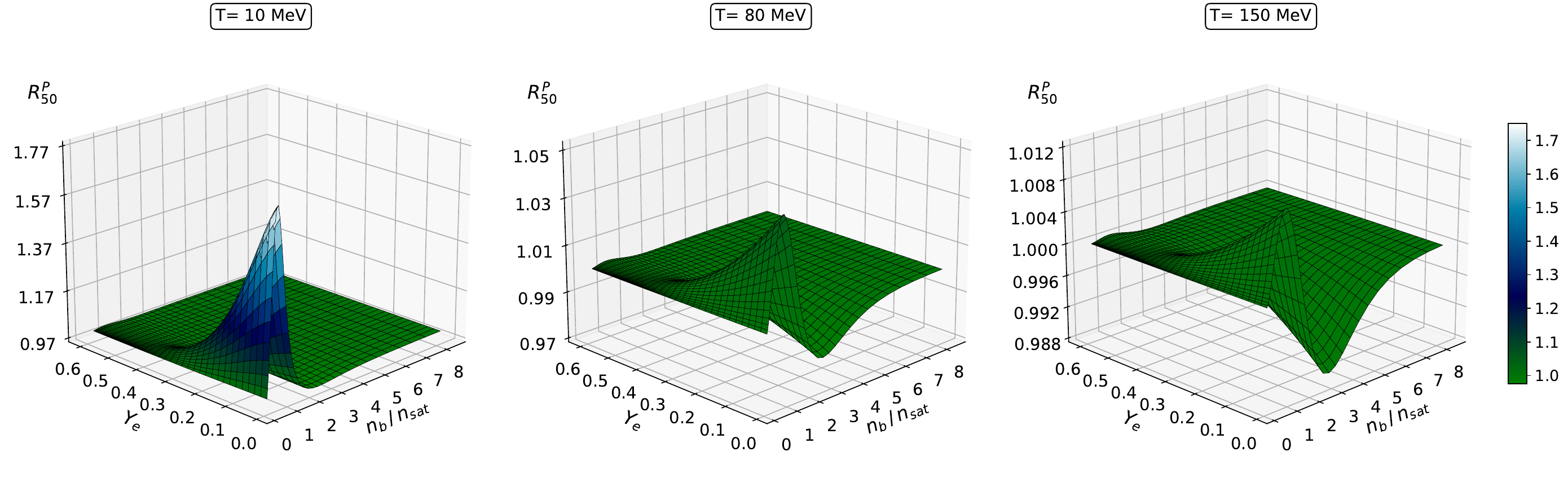}
\includegraphics[width=\linewidth,keepaspectratio]{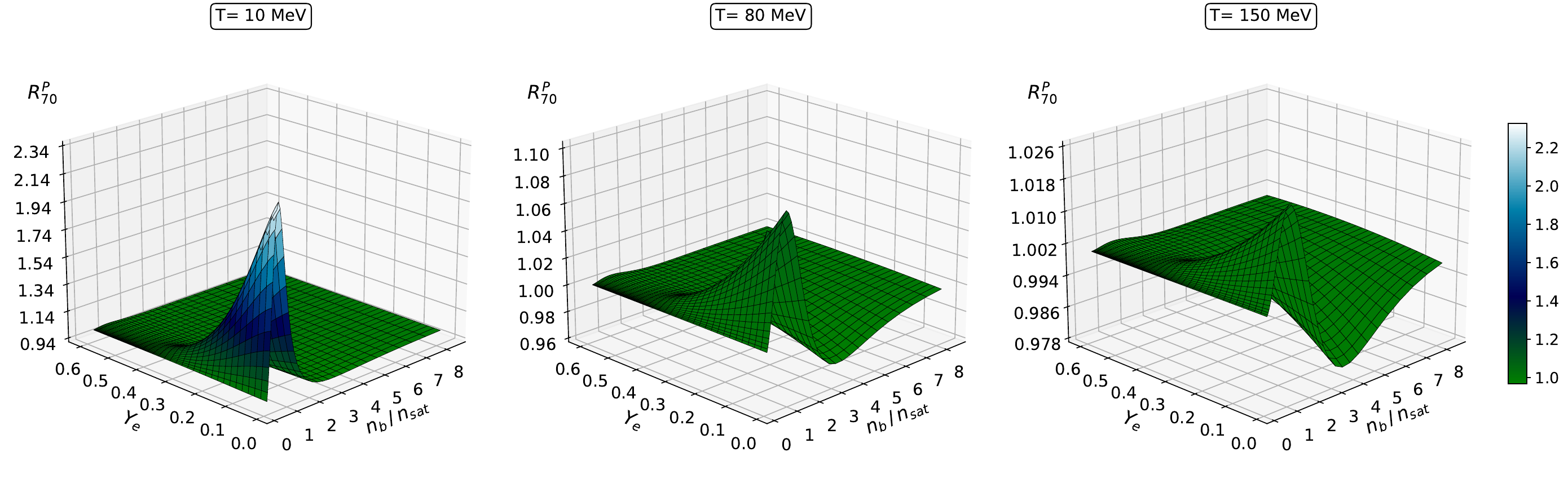}
\caption{Upper panel: Pressure of nucleonic matter as a function of baryon density (normalized by
  the saturation density $n_{\rm sat}$) and electron fraction for $\Lsym=30$~MeV and temperatures  $T=10$~MeV (left), $T=80$~MeV (middle) and $T=150$~MeV (right).
Middle panel: Ratio of the pressure of nucleonic matter with $\Lsym=50$~MeV  to the pressure of nucleonic matter with $\Lsym=30$~MeV  as a function of baryon density (normalized by
  the saturation density $n_{\rm sat}$) and electron fraction for temperatures $T=10$~MeV (left), $T=80$~MeV (middle) and $T=150$~MeV (right). Lower panel: Same as the middle panel but for the ratio of pressure at $\Lsym=70$ MeV to that at $\Lsym=30$ MeV.}
\label{fig:N_P_n_TConst} 
\end{center}
\end{figure*}
\begin{figure*}[tbh] 
\begin{center}
\includegraphics[width=\linewidth,keepaspectratio]{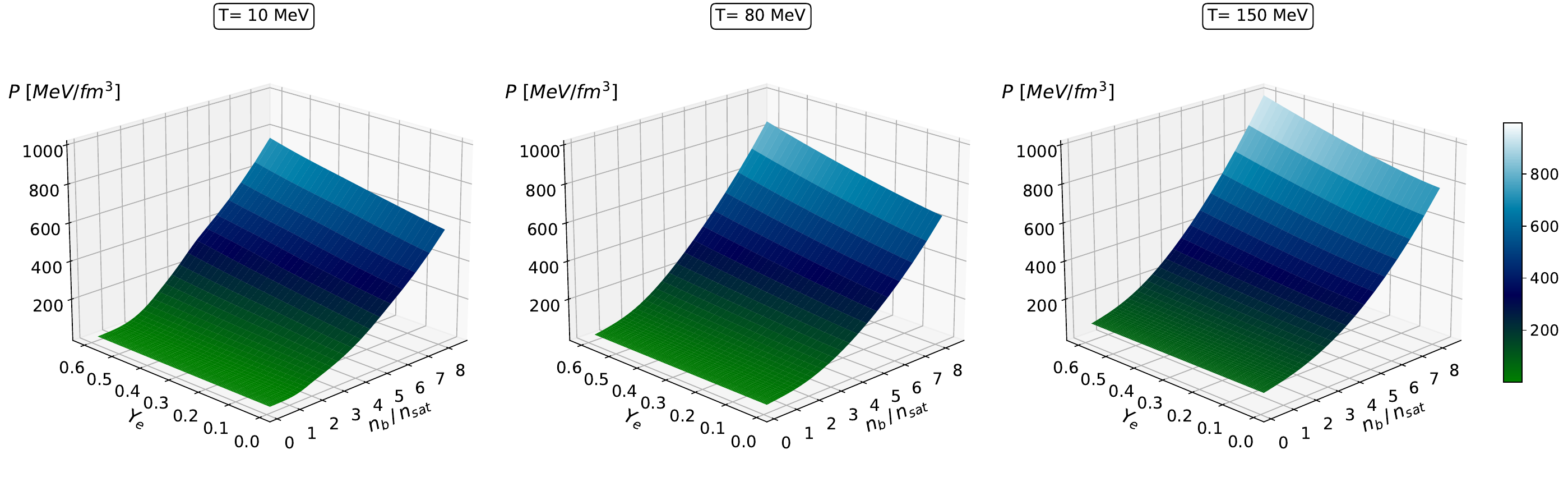}
\includegraphics[width=\linewidth,keepaspectratio]
{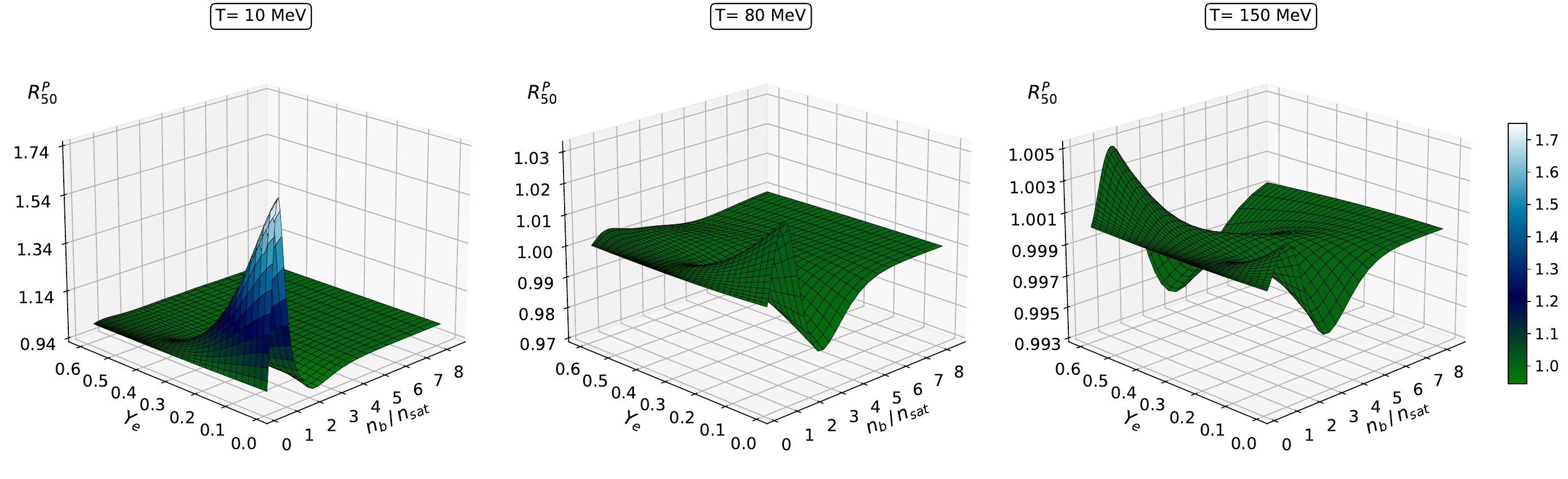}
\includegraphics[width=\linewidth,keepaspectratio]{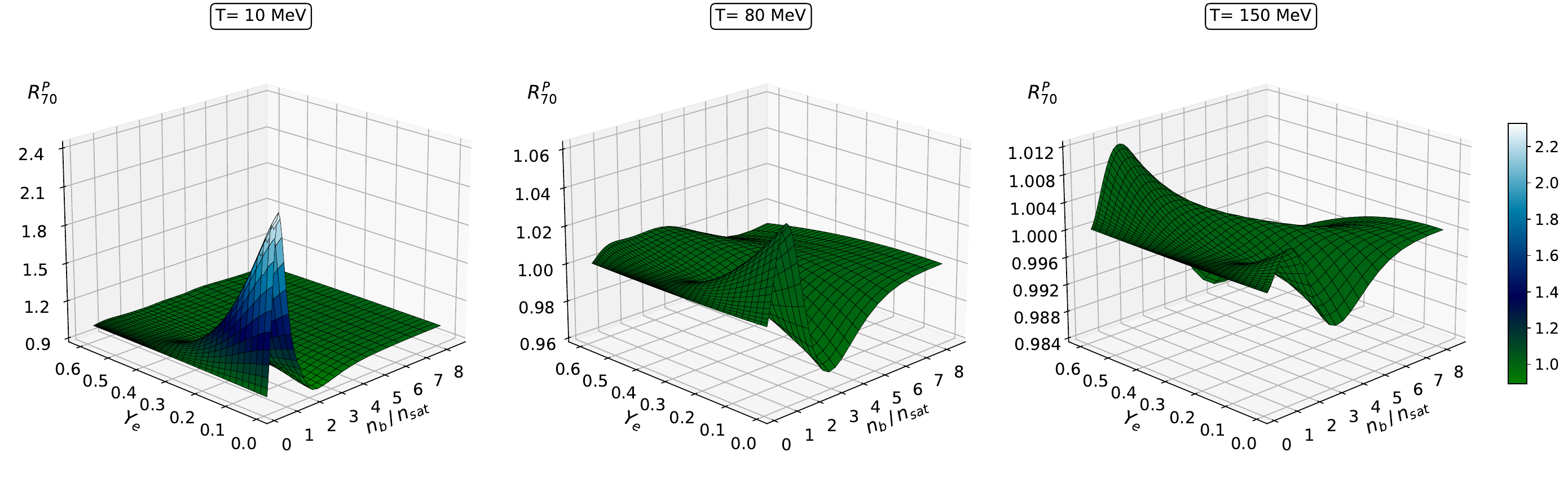}
\caption{
Upper panel: Pressure of hypernuclear matter as a function of baryon density (normalized by
  the saturation density $n_{\rm sat}$) and electron fraction for $\Lsym=30$~MeV and temperatures  $T=10$~MeV (left), $T=80$~MeV (middle) and $T=150$~MeV (right).
Middle panel: Ratio of the pressure of hypernuclear matter with $\Lsym=50$~MeV  to the pressure of hypernuclear matter with $\Lsym=30$~MeV as a function of baryon density (normalized by the saturation density $n_{\rm sat}$) and electron fraction for temperatures $T=10$~MeV (left), $T=80$~MeV (middle) and $T=150$~MeV (right). Lower panel: Same as the middle panel but for the ratio of pressure at $\Lsym=70$ MeV to that at $\Lsym=30$ MeV.
}
\label{fig:Y_P_n_TConst} 
\end{center}
\end{figure*}
\subsection{Matching to low-density matter for the general-purpose tables}
\label{subsec:Matching}

\begin{figure*}[!tbp]
\begin{center}
\includegraphics[width=\linewidth,keepaspectratio]{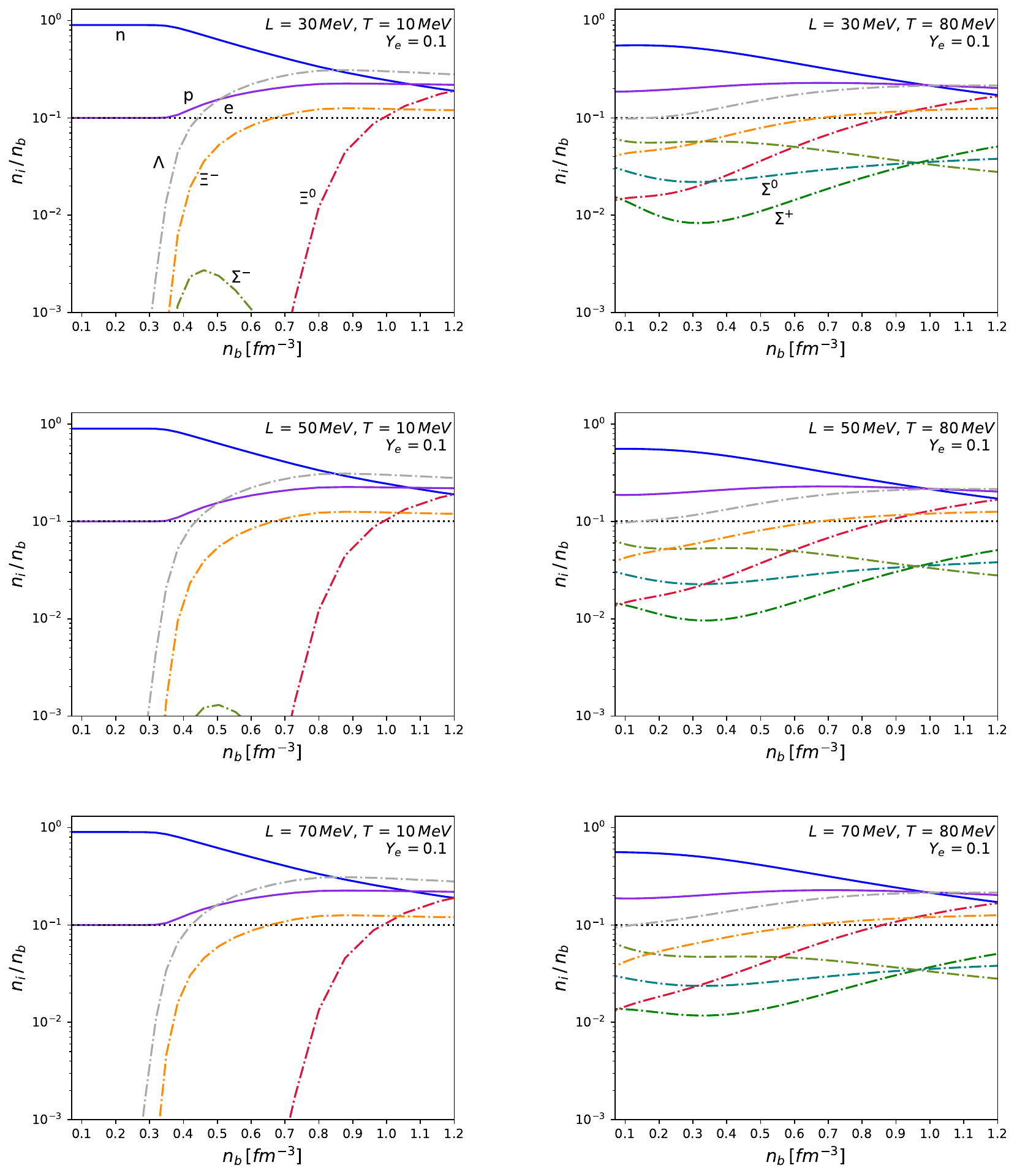}
\caption{Composition of matter for the three different values of $L$ at a constant electron fraction of $Y_{e} = 0.1$ and temperature of $T=10$~MeV (left panels) and $T=80$~MeV (right panels). All figures correspond to the hypernuclear EoS.}
\label{fig:abundances-T_const_Ye01}
\end{center}
\end{figure*}

\begin{figure*}[!tbp]
\begin{center}
\includegraphics[width=16cm,keepaspectratio]{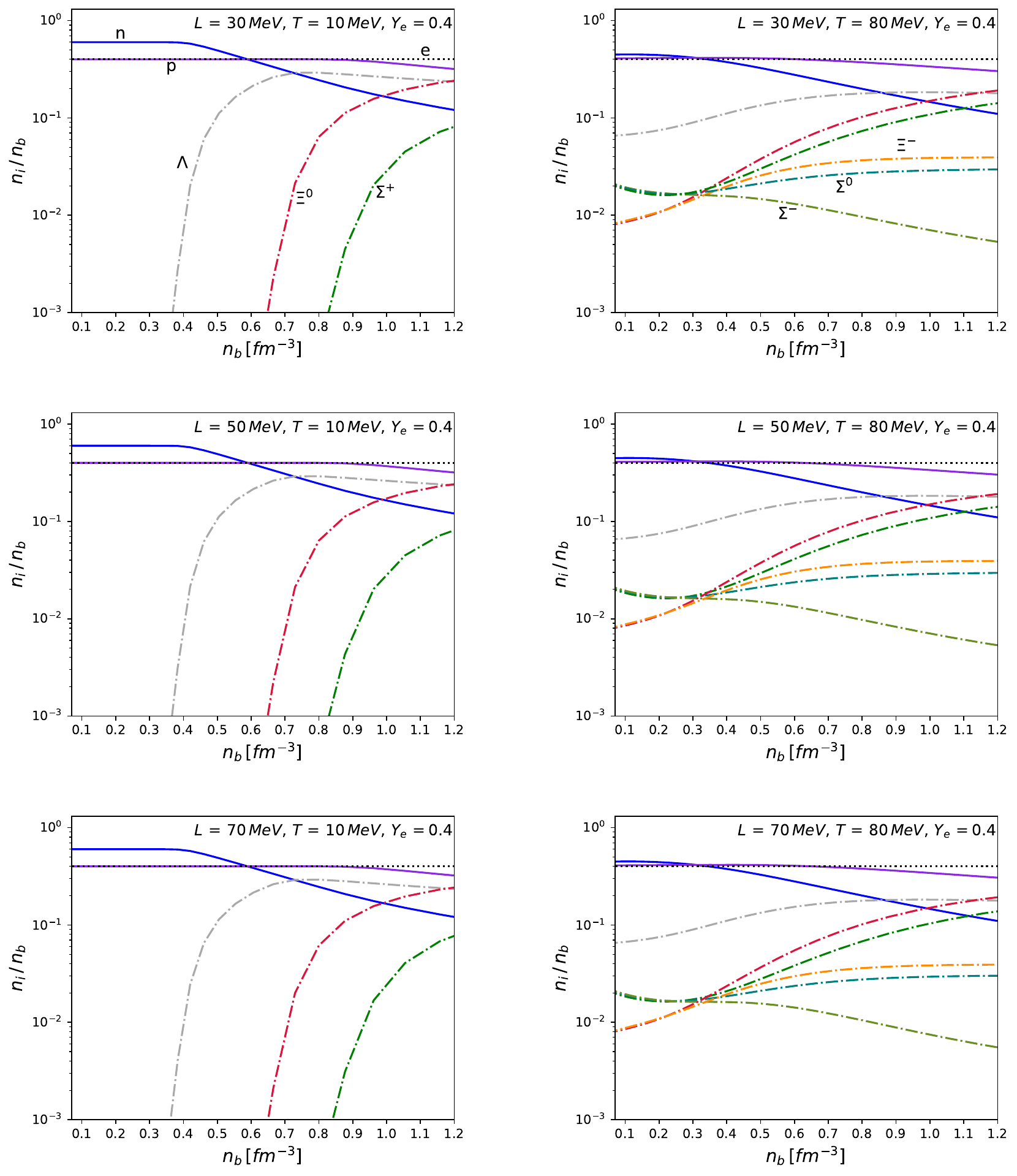}
\caption{Same as in Fig.~\ref{fig:abundances-T_const_Ye01}, but for an electron fraction of $Y_{e}=0.4$.}
\label{fig:abundances-T_const_Ye04}
\end{center}
\end{figure*}

In numerical simulations of CCSN or BNS mergers, in general, the EoS is incorporated in the form of tables covering the necessary ranges in thermodynamic parameters. The latter are mostly chosen to be temperature, baryon number density, and electron fraction (as mentioned above, muon fraction needs an additional evolution equation and is in general not included) with values typically in between $0.1 \lesssim T  \lesssim 100 $ MeV, $10^{-12} \lesssim n_B \lesssim 1 \mathrm{fm}^{-3}$, and $0.01 \lesssim Y_e \lesssim 0.6$~\cite{Oertel2017} to describe matter under the very different thermodynamic conditions occurring during the CCSN or the BNS merger. The models presented here only consider homogeneous matter, i.e. they are not adapted to the low-density and temperature region where nuclear clusters coexist with unbound nucleons. To extend our models into that region and produce a complete general-purpose table, we have chosen to match the high-density EoS to the low-density HS(DD2) one~\cite{Hempel2010}  at a density of $n_B = 0.04$ fm$^{-3}$. If for the given values of $T$ and $Y_e$, matter is not homogeneous in the original HS(DD2) table at this density, then the matching is performed at the lowest density for homogeneous matter. The HS(DD2) model describes inhomogeneous matter within an extended nuclear statistical equilibrium approach~\cite{Hempel2010}, treating the interaction of unbound nucleons within the DD2~\cite{Typel2010} model. At low densities, the DD2 interaction is very close to the models employed here and at the densities at which we perform the matching, only nucleons should be present. This ensures a smooth matching of inhomogeneous to homogeneous matter upon constructing our general-purpose table.  The only weakness is that we miss a fraction of hyperons and $\Delta$-resonances at high temperatures and low densities.  Indeed in this regime, the hyperon fractions are substantial, see Figs. 
\ref{fig:abundances-T_const_Ye01} and \ref{fig:abundances-T_const_Ye04}
and Refs.~\cite{Marques2017,Fortin2018,Oertel2017,Sedrakian2022,Sedrakian2023} and matching the purely nuclear HS(DD2) EoS within our tables induces small discontinuities in particular in the hyperon fractions. Their influence on thermodynamic quantities is small. We will improve on this point in a future version of the tables.

\section{Equation of state and composition of hypernuclear matter}
\label{sec:Num_EoS_Composition}

Our tables were generated through self-consistent solutions of the equations for the meson fields (in the static approximation) and the scalar and baryon densities \eqref{eq:density_b} and \eqref{eq:density_s} for fixed values of temperature,  density, and electron fraction. The neutrino contribution is typically added within the simulations from the employed neutrino treatment which also accounts consistently for neutrino trapping, which happens typically in dense matter above a temperature of several MeV. Muons are neglected likewise to limit the dimensionality of the table to three.

\begin{figure*}[t] 
\begin{center}
\includegraphics[width=\linewidth,keepaspectratio]{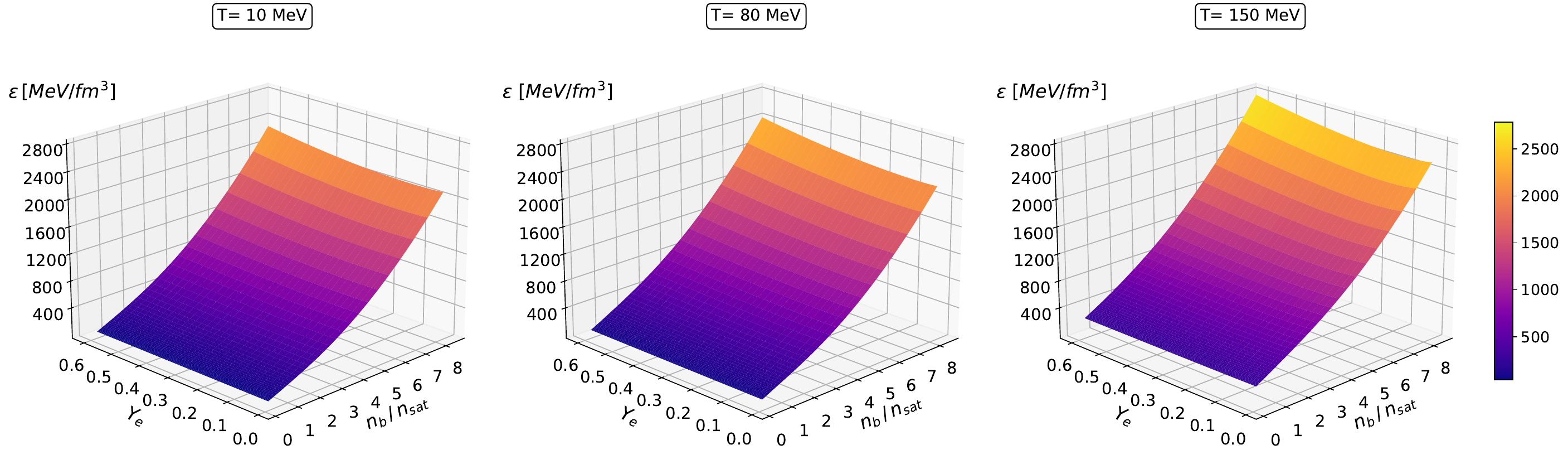}
\includegraphics[width=\linewidth,keepaspectratio]{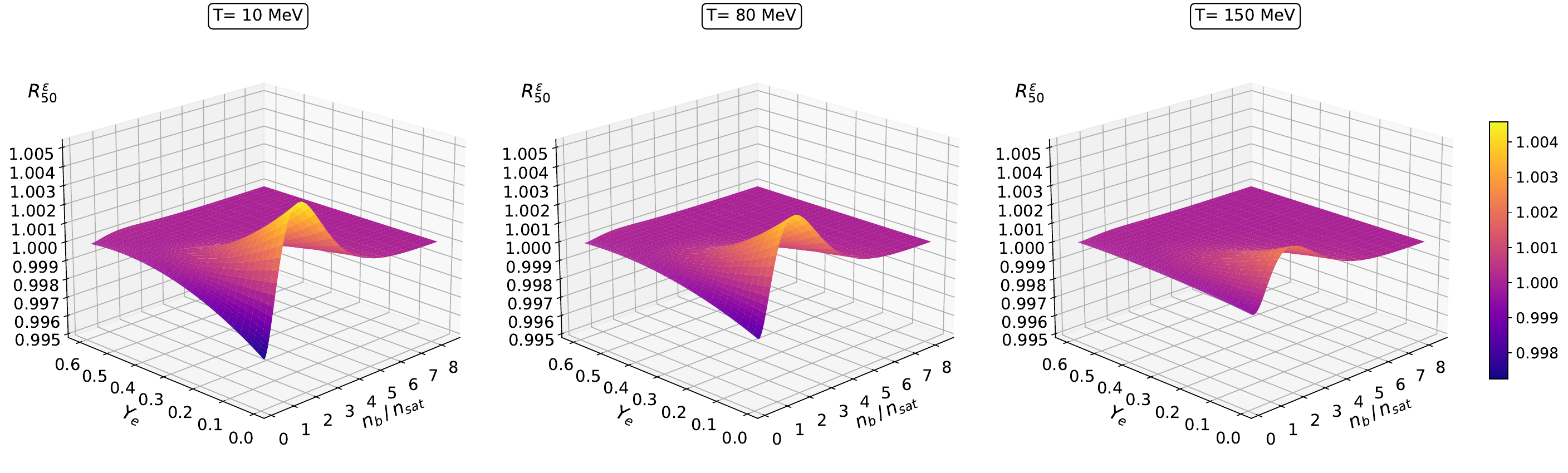}
\includegraphics[width=\linewidth,keepaspectratio]{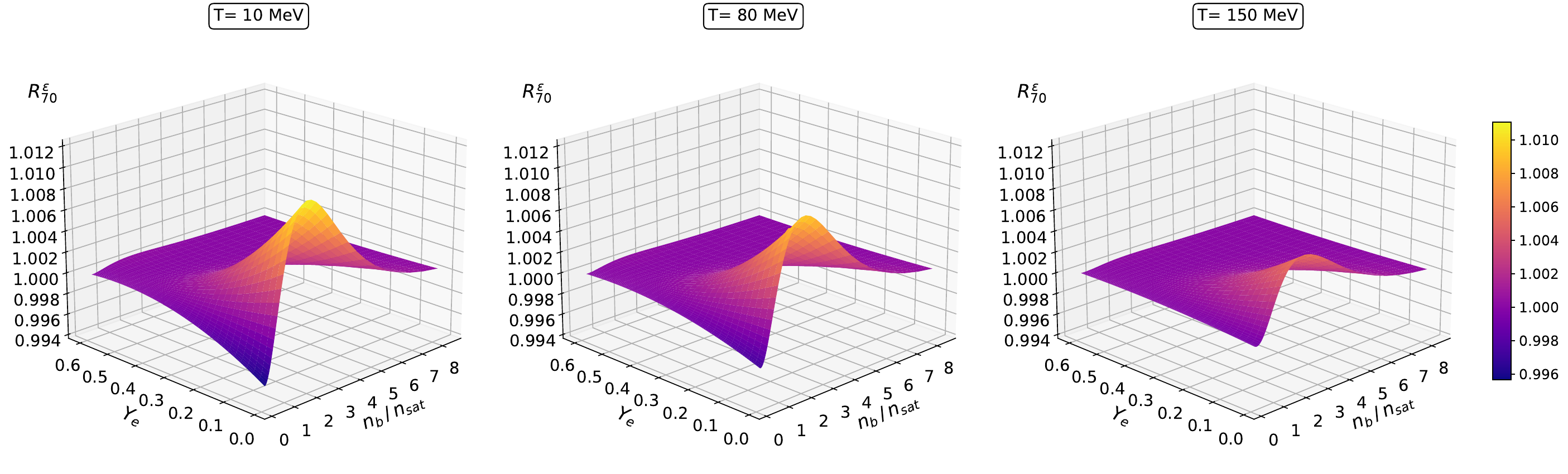}
\caption{
Upper panel: Energy density of nucleonic matter as a function of baryon density (normalized by
  the saturation density $n_{\rm sat}$) and electron fraction for $\Lsym=30$~MeV and temperatures  $T=10$~MeV (left), $T=80$~MeV (middle) and $T=150$~MeV (right).
Middle panel: Ratio of the energy density of nucleonic matter with $\Lsym=50$~MeV to energy density of nucleonic matter with $\Lsym=30$~MeV  as a function of baryon density (normalized by
  the saturation density $n_{\rm sat}$) and electron fraction for temperatures $T=10$~MeV (left), $T=80$~MeV (middle) and $T=150$~MeV (right). Lower panel: Same as the middle panel but for the ratio of energy density at $\Lsym=70$ MeV to that at $\Lsym=30$ MeV.
}
\label{fig:N_E_n_TConst} 
\end{center}
\end{figure*}
\begin{figure*}[t] 
\begin{center}
\includegraphics[width=\linewidth,keepaspectratio]{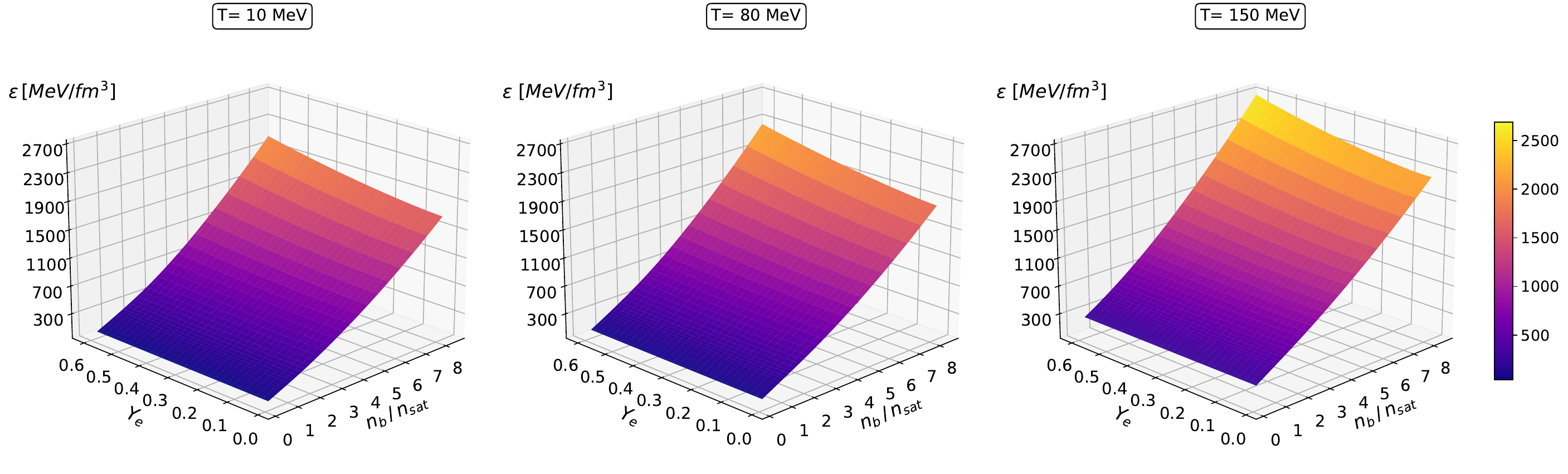}
\includegraphics[width=\linewidth,keepaspectratio]{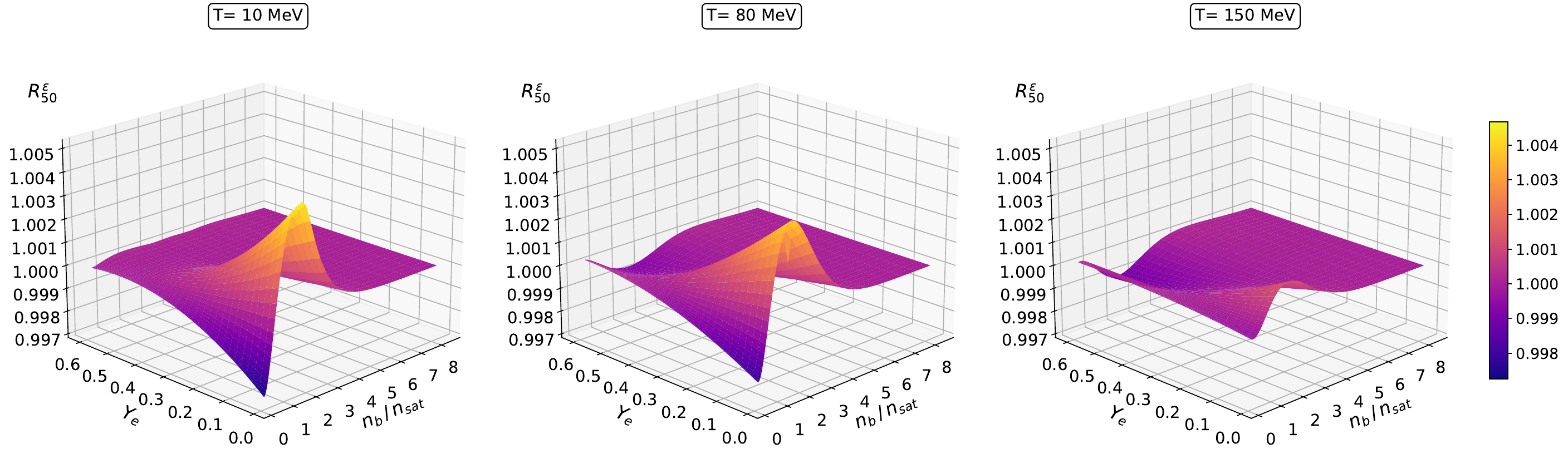}
\includegraphics[width=\linewidth,keepaspectratio]{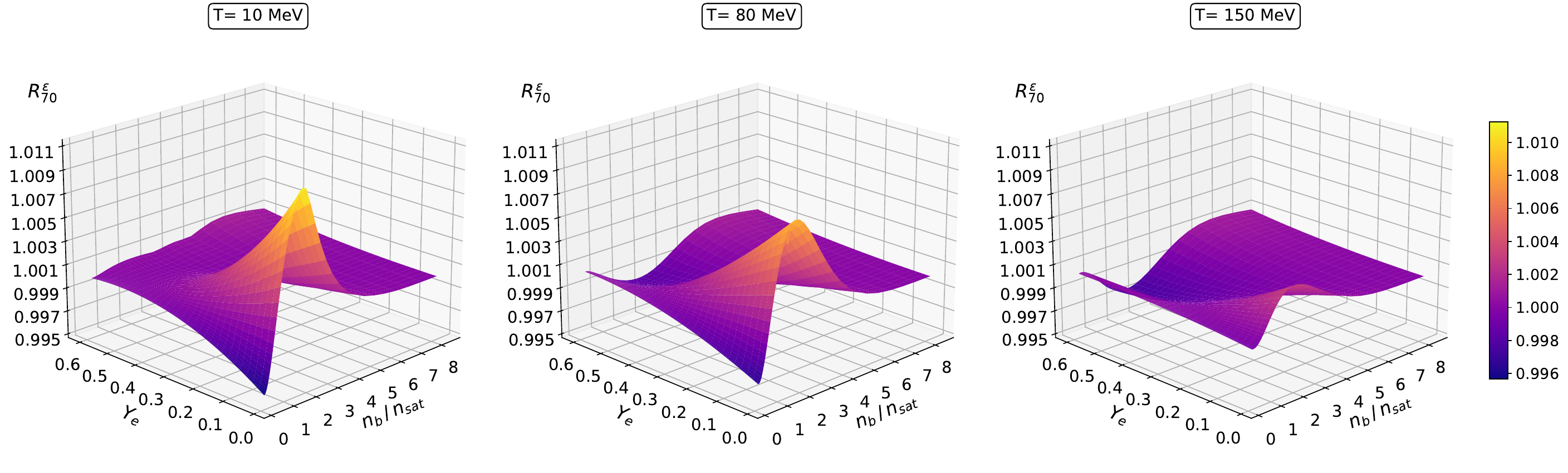}
\caption{
Upper panel: Energy density of hypernuclear matter as a function of  baryon density (normalized by
  the saturation density $n_{\rm sat}$) and electron fraction for $\Lsym=30$~MeV and temperatures  $T=10$~MeV (left), $T=80$~MeV (middle) and $T=150$~MeV (right).
Middle panel: Ratio of the energy density of hypernuclear matter with $\Lsym=50$~MeV  to the energy density of hypernuclear matter with $\Lsym=30$~MeV  as a function of baryon density (normalized by
  the saturation density $n_{\rm sat}$) and electron fraction for temperatures $T=10$~MeV (left), $T=80$~MeV (middle) and $T=150$~MeV (right). Lower panel: Same as the middle panel but for the ratio of energy density at $\Lsym=70$ MeV to that at $\Lsym=30$ MeV. 
}
\label{fig:Y_E_n_TConst} 
\end{center}
\end{figure*}
We next illustrate the content of the tables by showing selected results, which are obtained through cuts in three-dimensional space spanned by density, temperature, and electron fraction. The finite-temperature pressure as a function of density and electron fraction for nucleonic matter is shown in Fig.~\ref{fig:N_P_n_TConst} for  $\Lsym=30$ MeV and three values of fixed temperature. We show in the same figure 
 the ratios of pressures $P[\Lsym=50 {\rm~MeV}]/P[\Lsym=30 {\rm~MeV}]$ 
and $P[\Lsym=70  {\rm~MeV}]/P[\Lsym=30 {\rm~MeV}]$ to visualize the changes with the change in the symmetry energy slope $\Lsym$. As expected, pressure is an increasing function of density; The ratios of pressure for different $\Lsym$ values tend to unity in the limit of large densities where pressure is dominated by the value of $\Qsat$ and the influence of $\Lsym$ is negligible. The ratios increase as one moves away from the symmetric limit. It is also obvious that the ratios are numerically larger the larger the difference in $\Lsym$. Increasing the temperature for a fixed value of $\Lsym$ diminishes the ratios of the pressures as they are increasingly dominated by thermal effects rather than interactions. 

Figure~\ref{fig:Y_P_n_TConst} shows the EoS for the same parameters as in Fig.~\ref{fig:N_P_n_TConst}, but in the presence of hyperons. It is seen that this new feature strongly softens the EoS due to the onset of the new degrees of freedom, which allow for the reduction of the degeneracy pressure of nucleons.  The ratios of the pressures at various $\Lsym$ show the same trends seen in the nucleonic case since the factors that determine them are unchanged when hyperons are added.

Figure~\ref{fig:N_E_n_TConst} shows the energy density of nucleonic matter as a function of density and electron fraction for  $\Lsym=30$ MeV and again the same three values of fixed temperature. It also shows the ratios of energy densities ${\cal E}[\Lsym=50 {\rm~MeV}]/{\cal E}[\Lsym=30 {\rm~MeV}]$ 
and ${\cal E}[\Lsym=70 {\rm~MeV}]/{\cal E}[\Lsym=30 {\rm~MeV}]$.  The general features that we already discussed in the case of pressure are repeated in the case of energy density as well.  (a) It is an increasing function of density and has a minimum at the isospin symmetric limit.  (b) The ratios of energy densities for different $\Lsym$ values are peaked at $Y_e\to 0$ (pure neutron matter) as the asymmetry energy is maximal in this limit. (c) Again, as in the case of pressure ratios, the ratios of energy densities are numerically larger the larger the difference in $\Lsym$. 

Figure~\ref{fig:Y_E_n_TConst} shows the same quantities as in Fig.~\ref{fig:N_E_n_TConst} but for hypernuclear matter. The energy density ratios at different $\Lsym$ values exhibit the same trends already observed in the nucleonic case, and we do not repeat the discussion here.


\subsection{Composition of matter}
\label{subsec:Composition}

We next turn to the composition of matter under the conditions considered in this work. Figure \ref{fig:abundances-T_const_Ye01} shows the composition of finite-temperature hyperonic matter at two temperatures $T=10$ and 80~MeV for three values of $\Lsym=30$, 50 and 70 MeV and fixed $Y_e=0.1$.  
At  $T=10$~MeV
hyperons $\Lambda$,  $\Xi^-$, $\Sigma^-$ and $\Xi^0$ appear in the given order with increasing density, with the $\Sigma^-$ hyperon fraction being strongly suppressed by the highly repulsive potential in nuclear matter at saturation density. Since the chemical potentials are still much larger than the temperature, this arrangement is qualitatively the same as the one at zero temperature and shows also relatively sharp thresholds for the appearance of hyperons. The large negative charge chemical potential in matter with low electron fractions favors here the negatively charged hyperons over their isospin partners, see e.g. the discussion in~\cite{Oertel2016}. For higher temperature  $T=80$~MeV the thresholds disappear and hyperon abundances extend deep in the low-density regime. The isospin triplet of $\Sigma^{\pm,0}$ is now thermally supported with amounts comparable to other hyperons. 
As pointed out in Ref.~\cite{Sedrakian2021,Sedrakian2022} there is a special isospin degeneracy point where the fractions within each isospin multiplet coincide. This is visible for $\Sigma$ and $\Xi$ hyperons at $T=80$~MeV. At this degeneracy point, there is a reversal in the dominance of the abundances of the hyperons present. For example, as density increases, the $\Sigma^-$  hyperon goes over from being the most abundant to the least abundant hyperon in the $\Sigma$-hyperon multiplet. Similar behavior is seen as well for $\Xi^-$ and $\Xi^0$ fractions. The mechanism underlying the isospin degeneracy point is discussed in Ref.~\cite{Sedrakian2022},
where it is pointed out that this effect is related to the vanishing of the charge chemical potential at that point. The variations of the abundances of hyperons with the value of $\Lsym$ are large quantitatively, see however the visible suppression of $\Sigma^-$ hyperons with increasing $\Lsym$. 

Figure \ref{fig:abundances-T_const_Ye04} shows the same for an electron fraction of $Y_e = 0.4$. The higher charge chemical potential in this less neutron-rich regime leads to a different composition where the negatively charged particles within an isospin multiplet are no longer favored. This can be seen at $T = 10$ MeV with the $\Xi^0$ and the $\Sigma^+$ appearing first and at $T = 80$ MeV, the isospin degeneracy point is shifted to lower baryon number densities.  The value of $\Lsym$ only has a minor influence on the results because the contribution of the symmetry energy is small for nearly symmetrical matter.

\section{Cold equation of state and astrophysical constraints}
\label{sec:MR}

\begin{figure}[tbh] 
\begin{center}
\includegraphics[width=\linewidth,keepaspectratio]{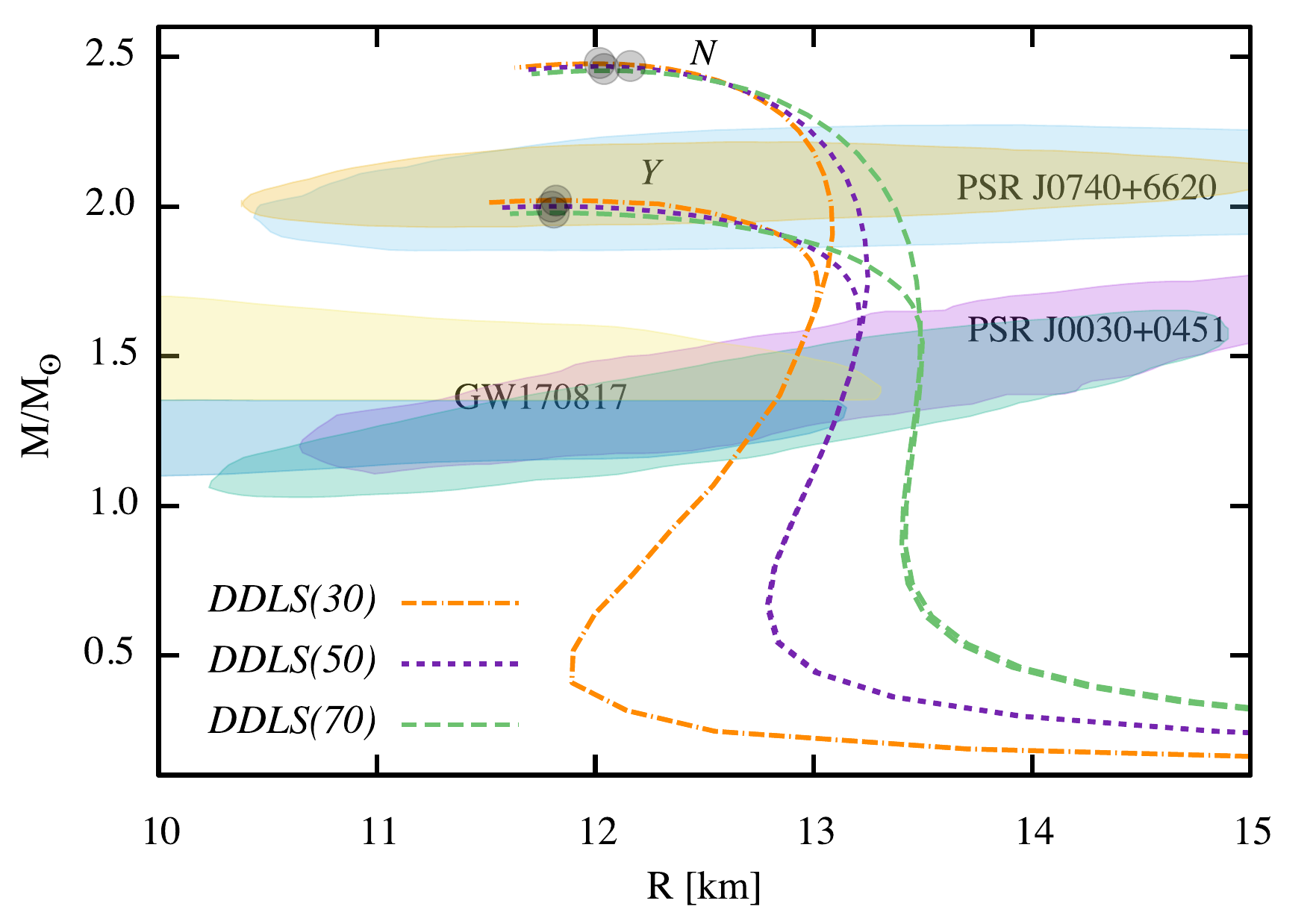}
\includegraphics[width=\linewidth,keepaspectratio]{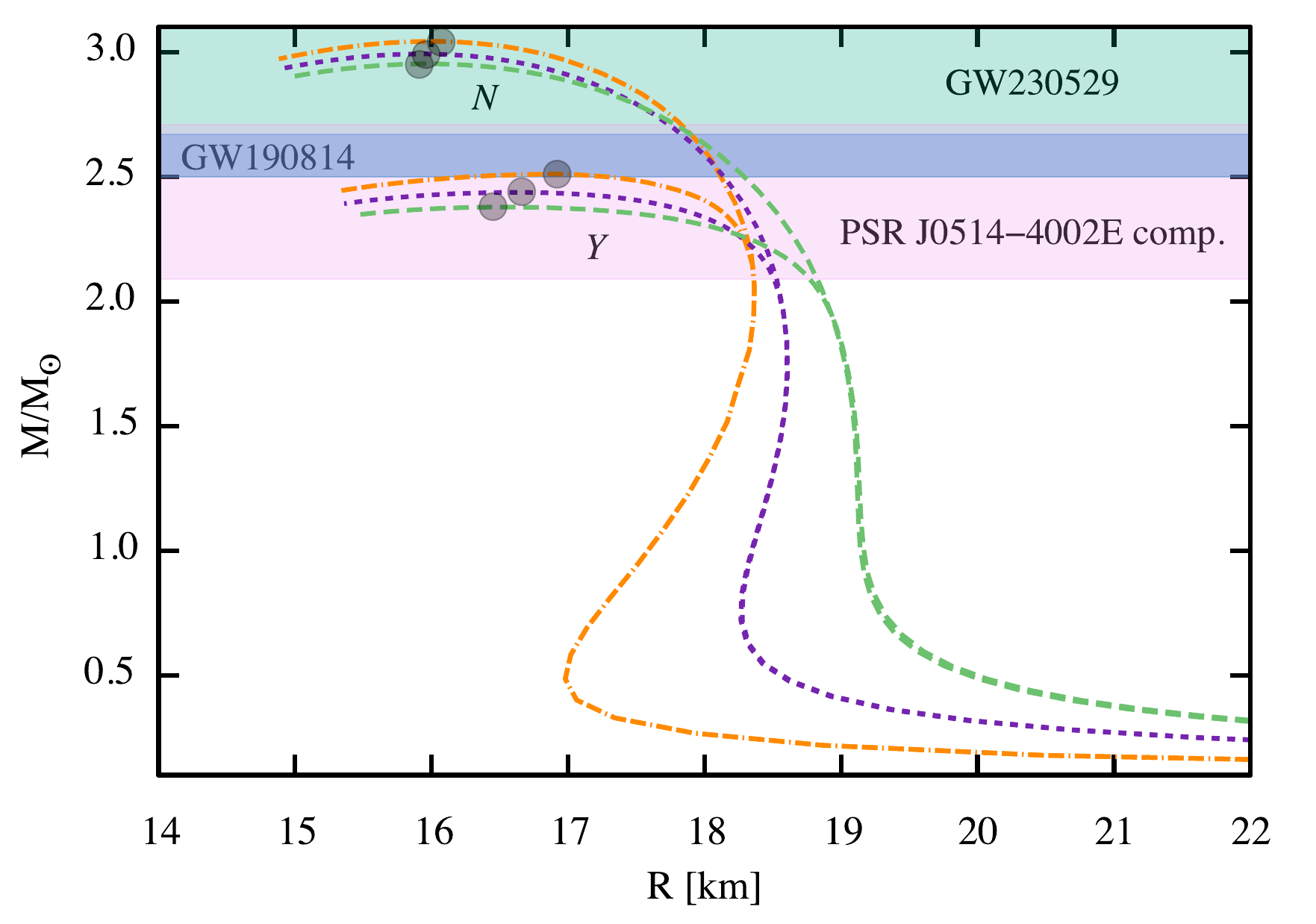}
\caption{Gravitational mass versus radius for non-rotating (upper
  panel) and maximally fast rotating (lower panel) stars.  In the case
  of rotating stars, $R$ refers to the equatorial radius. The upper
  three branches, labeled as $N$, correspond to nucleonic, and the
  lower three labeled as $Y$ -- to
  hyperonic stars for EoS models DDLS(30), DDLS(50) and DDLS(70) for fixed
  value of $\Qsat =400$ MeV. The ellipses in the upper panel show
  90\% CI regions for PSR J0030+0451~\cite{Riley2019,Miller2019}, PSR J0740+6620~\cite{Riley2021,Miller2021} and
  gravitational wave event GW170817~\cite{Abbott2017}. The colored regions correspond to the light member in the GW190814 event (blue)~\cite{LVC_GW190814}, the heavy member in the GW230529 event (light green)~\cite{LVC_GW230529}, and the heavy companion of PSR J0514-4002E (violet)~\cite{Barr2024}. The dots indicate the location of the maximum mass for any given sequence, see also Table~\ref{tab:NS}.}
\label{fig:MR} 
\end{center}
\end{figure}

While the finite-temperature EoS is needed for transient astrophysical
scenarios involving neutron stars, the cold (zero-temperature) limit
of the EoS is sufficient to describe their secular time scale
evolution independent of their observational manifestations as
pulsars, accreting X-ray neutron stars, etc. Currently, available
astrophysical constraints on the global parameters of neutron stars
put limits on the cold $\beta$-equilibrated EoS, therefore we complete the discussion of
the finite-temperature EoS by confronting its zero temperature limit
with the observations, see also  
Ref.~\cite{Li2023}. 
In Fig.~\ref{fig:MR} the 90\% CI  ellipses show
three key constraints involving PSR J0030+0451~\cite{Riley2019,Miller2019}, PSR J0740\-+6620~\cite{Riley2021,Miller2021}, and the
gravitational wave event GW170817~\cite{Abbott2017}. The first two objects are pulsars with constrained radii.  
PSR  J0030+0451 has 
a gravitational mass $M=1.34^{+0.15}_{-0.16} M_{\odot}$ and radius
$R= 12.71^{+1.14}_{-1.19}$~km (68\% CI)~\cite{Riley2019}. 
An alternative evaluation leads to 
$M = 1.44^{+0.15}_{-0.14}\, M_{\odot}$ and
$R = 13.02^{+1.24}_{^-1.06}$~km (68\% CI)~\cite{Miller2019}.
For the more massive compact star PSR J0740+6620
one finds the mass $2.08\pm 0.07 M_{\odot}$ and 
radius $13.7^{+2.6}_{-1.5}$~km~\cite{Miller2021} or, alternatively,  $2.072^{+0.067}_{-0.066} M_{\odot}$ and 
$12.39^{+1.30}_{-0.98}$~km (68\% CI)~\cite{Riley2021}.
\begin{table}
 \centering
    \setlength{\tabcolsep}{1pt} 
  \begin{tabular}{ccccccc}
    \hline\hline
   {\rm Model} & $M_{G,{\rm max}}$ &$R_{\rm max}$ & $\varepsilon_{B,{\rm max}}$
    & $R_{1.4M_{\odot}}$ &  $\Omega^K_{\rm max}/10^4$ \\
    &     $[M_{\odot}]$ &   [km]   &   [$10^{15}$\,g\, cm$^{-3}$]  &  [km]  & [Hz] \\
    \hline
 $N$, DDLS(30)          & 2.48  & 12.02  &    1.85     &  12.87&  - \\
 $N$, DDLS(50)          & 2.47  & 12.16  &    1.90     &  13.15 &  -\\
 $N$, DDLS(70)          & 2.46  & 12.04  &    1.85     &  13.47 &  - \\
 $Y$, DDLS(30)          & 2.02  &  11.82  &    1.94    &  12.87 & - \\
 $Y$, DDLS(50)          & 2.00  &  11.80  &    2.02    &  13.15 & -\\
 $Y$, DDLS(70)          & 1.98  &  11.81 &     2.08    &  13.48 & -\\
    \hline
    \hline
 $N$, DDLS(30)          & 3.04  &  16.07    &    1.53     &  18.06& 0.96 \\
 $N$, DDLS(50)          & 2.99  &  15.96    &    1.60     &  18.53& 0.97  \\
 $N$, DDLS(70)          & 2.95  &  15.91    &    1.67     &  19.10& 0.97 \\
 $Y$, DDLS(30)          & 2.51  &   16.92    &    1.40    &   18.06 &  0.82 \\
 $Y$, DDLS(50)          & 2.44  &   16.66    &    1.53    &   18.53 &  0.83  \\
 $Y$, DDLS(70)          & 2.38  &   16.45    &    1.68    &   19.10 &  0.83 \\
    \hline
    \hline
  \end{tabular}
  \caption{ The upper half of the table refers to the properties of
    non-rotating spherically symmetric cold $\beta$-equilibrated,
    neutrino-transparent, compact stars based on the EoS models
    considered in this work. The first three columns show the maximum
    gravitational mass ($M_{G,{\rm max}}$), the corresponding radius
    $(R_{\rm max})$ and central energy density
    ($\varepsilon_{B,{\rm max}}$) for the (cold) EoS with nucleons
    only ($N$) and hyperons ($Y$).  The remaining column shows the
    radius at $M_G = 1.4M_{\odot}$ mass. The lower half shows the same
    quantities for maximally rotating (Keplerian) sequences. 
    The listed radii correspond to equatorial ones in this case. 
    In addition, we show the Keplerian frequency $\Omega^K_{\rm max}$ for maximum mass
    nucleonic and hyperonic stars. The same values for a $1.4M_{\odot}$ stars, which coincide in the cases of nucleonic and hyperonic stars, 
    are given by $\Omega^K_{1.4M_{\odot}}/(10^4\, {\rm Hz}) =0.57, \, 0.54,\, 0.52$ for DDLS(30), DDLS(50) and DDLS(70) models. 
    }
  \label{tab:NS}
\end{table}

The static solutions of Einstein's equations in spherical symmetry
were obtained by solving the Tolman-Oppen\-heimer-Volkoff
equations~\cite{Oppenheimer1939} for cold $\beta$-equilibrated EoS models DDLS(30), DDLS(50)
and DDLS(70) for purely nucleonic matter (labeled as $N$) and
hypernuclear matter (labeled as $Y$), see Fig.~\ref{fig:MR}.  In
addition, the same figure shows the $M$-$R$ relations for maximal
fast rotating (Keplerian) sequences for rigid rotation 
for the same EoSs. These were
computed with the RNS code~\cite{Stergioulas1995}.  
Table \ref{tab:NS} lists the
maximal gravitational mass $M_{G,\rm max}$ of each sequence
considered, as well as the corresponding radius $(R_{\rm max})$, and
central density ($\varepsilon_{B,{\rm max}}$). The well-known
softening of the EoS once hyperons are allowed results in the lower
maximum masses of non-rotating and rapidly rotating stars, see
Fig.~\ref{fig:MR} and Table~\ref{tab:NS}.  The masses and radii of the
non-rotating sequences are compatible with the NICER inferences for
canonical (i.e. $M\sim 1.4M_{\odot}$) and massive (i.e.
$M\sim 2M_{\odot}$) compact stars.  It is evident that the $N$ and $Y$
sequences differ only when the central density of a configuration is
above the threshold for the onset of hyperons.  The $M$-$R$ tracks are
fully consistent with GW170817 ellipses for DDLS(30) and DDLS(50) models
but require a smaller radius than predicted by the DDLS(70) model. We still
keep this model in our collection as we aim to cover a broad range
of $\Lsym$ values. Note that CDF models that allow for $\Delta$ resonances
in addition to hyperons can produce smaller radii for intermediate-mass stars without affecting the maximum mass of a sequence, see Ref.~\cite{Schurhoff2010,Drago2014,Cai2015,Raduta2020,Malfatti2020,Raduta2022}.

The lower panel of Fig.~\ref{fig:MR} shows the mass-radius diagram for maximally fast rotating nucleonic and hyperonic stars, where the radius is the equatorial one. Table~\ref{tab:NS} lists the key parameters of maximally fast rotating stars - the gravitational mass, equatorial radius, and Keplerian frequency for the maximum mass star. The interest in rotating hypernuclear (and $\Delta$-admixed) compact stars arose in connection with the possibility that the light companion in the highly asymmetric binary compact object coalescence event GW190814~\cite{LVC_GW190814} with estimated mass in the range $2.5\le M/M_{\odot}\le 2.67$ is a rotating hypernuclear star. The range of inferred masses is within the "mass gap" where neither neutron stars nor black holes were found and are also hard to form with current theoretical models. This scenario has been explored with the DDME2 parametrization with some variations of the hyperonic coupling constants~\cite{Sedrakian2020,Li2020,Dexheimer2021b,RuiFu2022}.  Ref.~\cite{RuiFu2022}
finds hypernuclear stars with masses close to 2.5$M_{\odot}$ can be achieved  in the case of maximally fast (Keplerian) rotation, but their values of $\Qsat$ 
are much larger than those considered here.  It was concluded that the GW190814 event is likely to be a low-mass black hole rather than a supramassive neutron star. Here we confirm the conclusion reached earlier that hyperonization 
precludes highly massive hypernuclear stars. The lower panel of Fig.~\ref{fig:MR} 
also shows two additional candidates for the mass-gap objects: the GW230529 event where the more massive object has an inferred mass in the range $2.5\le M/M_{\odot}\le 4.5$~\cite{LVC_GW230529} and PSRJ0514-4002E with the mass range 
of the primary $2.09\le M/M_{\odot}\le 2.71$~\cite{Barr2024}, the heavy member in the GW230529 event (light green), and the heavy companion of PSR J0514-4002E (violet). Since the lower limit for GW230529 coincides with errors with the GW190814, our conclusions apply to this object as well. The lower limit on the mass of the companion of  PSRJ0514-4002E allows for a hypernuclear neutron star even without rotation. 
\begin{figure}[!tbp] 
\begin{center}
\includegraphics[width=\linewidth,keepaspectratio]{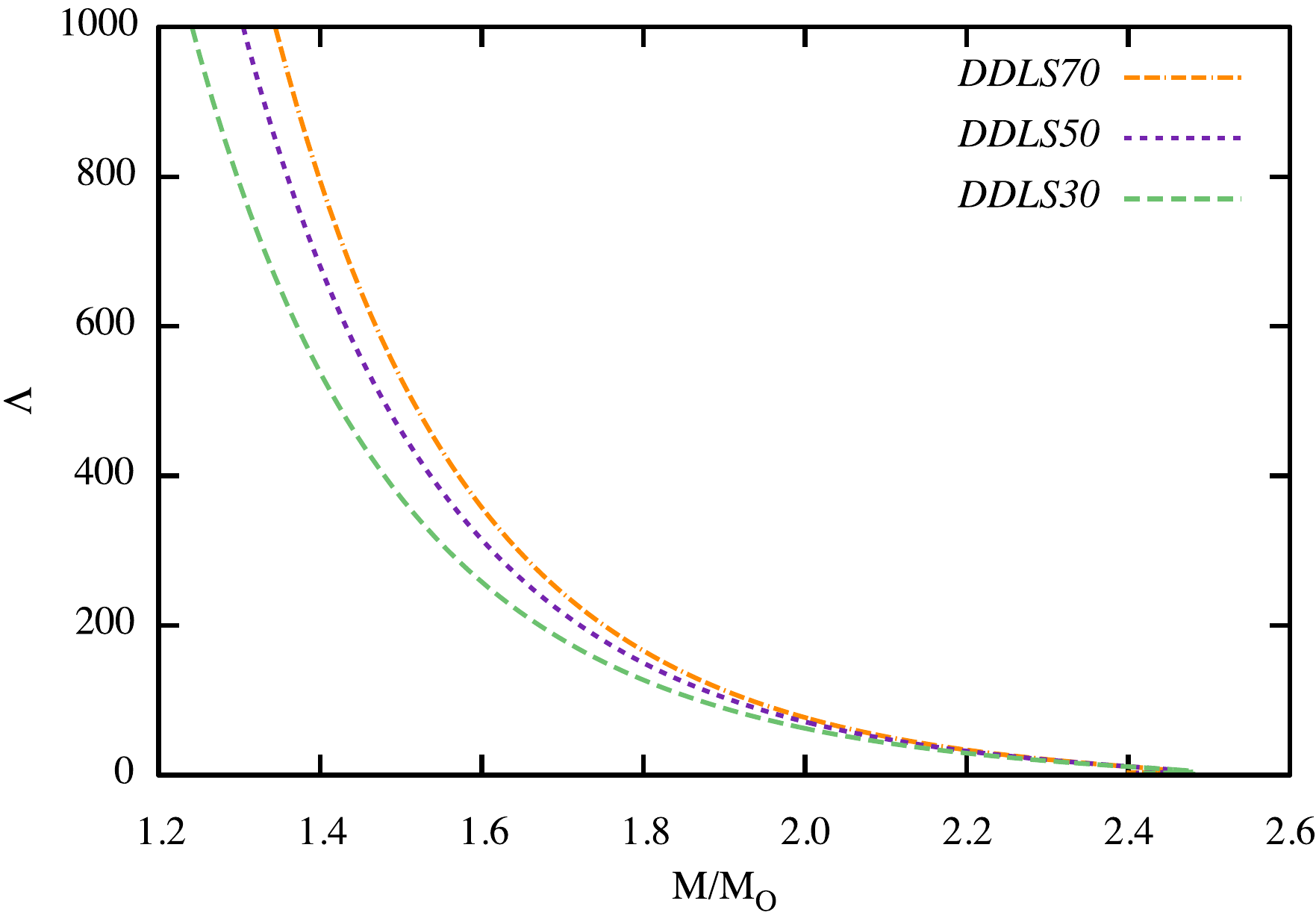}
\caption{Tidal deformability $\Lambda$ of nucleonic models DDLS30, DDLS50, and DDLS70. Current constraint, inferred from GW170817 event, sets an upper limit on tidal deformability  for an $1.4M_{\odot}$ mass star $\Lambda_{1.4}< 800$.
 Data taken from Ref.~\cite{Li2023}.}
\label{fig:ML} 
\end{center}
\end{figure}

In addition to the constraints on the mass-radius diagram, the dimensionless tidal deformability $\Lambda$ inference for the GW170817 offers an additional constraint on the EoS of dense matter. For the nucleonic models adopted here, the deformabilities were computed in Ref.~\cite{Li2023} and are shown in Fig.~\ref{fig:ML}.  The GW170817 event sets an upper limit on tidal deformability $\Lambda_{1.4}< 800$ for an $1.4M_{\odot}$
star~\cite{Abbott2017}. The deformability computed for the present models 
exceeds this limit for $M\le 1.39M_{\odot}$ in the case 
of DDLS70 model,  $M\le 1.34M_{\odot}$ for DDLS50 model 
and  $M\le 1.30M_{\odot}$ for the DDLS30 model, i.e., for current models the 
tidal deformability of a $1.4M_{\odot}$ mass 
star is consistent with the bound above. 
The hyperonic sequences branch off from the nucleonic ones when the mass is above $M\simeq 1.6M_{\odot}$ - the upper range of the mass of one of the stars involved in GW170817. Therefore, constraints on hyperonic sequences are the same as for the nucleonic ones. Nevertheless, we note that hyperonic models are softer than their nucleonic counterparts and consequently their deformabilities are smaller than the nucleonic ones for stars that have central densities above the hyperon onset, i.e. a high enough mass, see e.g.~Ref.~\cite{Li2019ApJ} for more details.

\begin{figure*}[!tbp] 
\begin{center}
\includegraphics[width=0.9\linewidth,keepaspectratio]{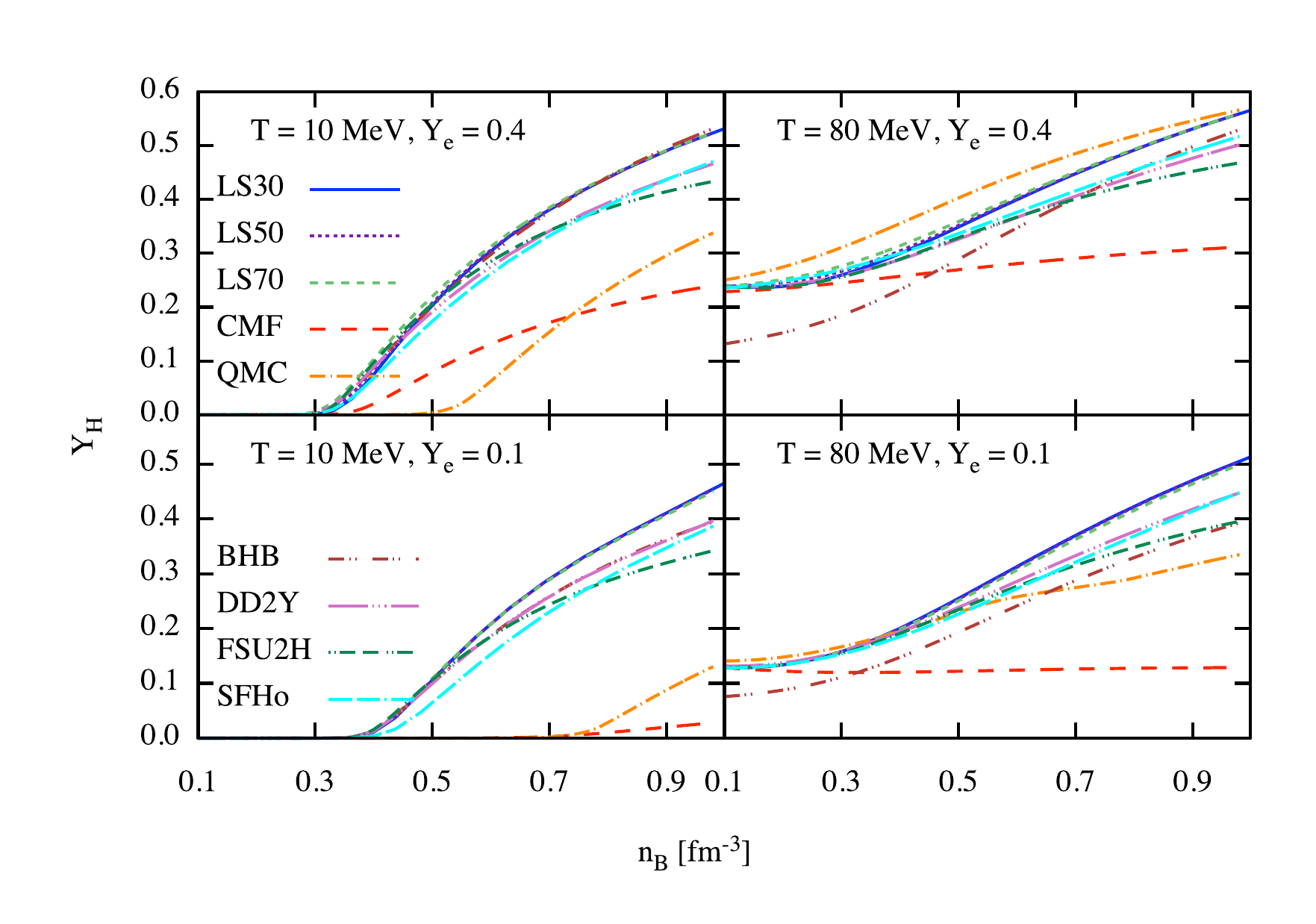}
\caption{ A comparison of hyperon fraction $Y_H$  as a function of
  baryon density for various fixed values of temperature $T$ and
  electron fraction $Y_e$ as predicted by different EoS models. The
  labelings of various curves are explained in the text.
}
\label{fig:hyperonfraction} 
\end{center}
\end{figure*}

\section{Comparison with previous work}
\label{sec:Comparison}
We complement now the discussion of our results with a brief
comparison of the models used in this work with those that exist in the
literature and are available on {\sc CompOSE} repository. Our focus is
on the general-purpose EoS, i.e., those that cover the
density-temperature and electron fraction parameter space. In addition, we will restrict our discussion to those models that include hyperons and
at the same time allow for a cold maximum TOV mass above the current
observational lower limit $M\gtrsim 2M_{\odot}$.  These restrictions
reduce the number of alternatives further.
\begin{table}
\begin{tabular}{lcccccc}
\hline 
 Model & $n_{\text {sat }}$ & $-E_B$ & $K$ & $E_{\text{sym}}$ & $L$  &
                                                                       Ref.\\
            &  [fm$^{-3}$]    & [MeV]   &[MeV]  & [MeV]               & [MeV]  &  \\
\hline
 \text {DD2} & 0.149     & 16.0 & 243 & 31.7 & 55.0 & \cite{Typel2010}\\
 \text {DDME2} & 0.152 & 16.14 & 251 &  32.3 & 51.3 & \cite{Li2019}\\
  \text {DDLS(30)} & 0.152 & 16.14 & 251 &  30.1 & 30 & \cite{Li2023}\\
   \text {DDLS(50)} & 0.152 & 16.14 & 251 &  32.2  & 50 & \cite{Li2023}\\
    \text {DDLS(70)} & 0.152 & 16.14 & 251 &  34.0 & 70 & \cite{Li2023}\\
 \hline
\text{QMC}      & 0.156  & 16.2 & 292 &  28.5 & 54 & \cite{Stone2021}\\
\text{CMF}    & 0.150   & 16.0 & 300 &  30.0 & 88 & \cite{Dexheimer2015}\\
\hline
 \text {SFHo} & 0.158 & 16.2 & 245 & 31.6 & 47.1 & \cite{Steiner2013}\\    
  \text {FSU2H} & 0.150 & 16.28 & 238 & 30.2 & 41.0 & \cite{Tolos2017}\\  
\hline
\end{tabular}
\caption{Nuclear characteristics at saturation density 
of models for which general-purpose EoS for hypernuclear matter
are available on {\sc CompOSE} database. Here $E_B$ is the binding energy per particle in symmetrical nuclear matter at saturation density $n_{\text {sat }}$ and $K$ is the compressibility. Note that the symmetry energy 
$E_{\text {sym }}$ of models DDLS(30), DDLS(50) and DDLS(70)  is fixed 
at the crossing-density 0.110 fm$^{-3}$ to value 
$E_{\text{sym}}=27.09$ MeV corresponding to the one
predicted by the DDME2 functional. Consequently,  
the symmetry energy and its slope coefficient $L$ 
at saturation density are changing in a correlated manner.
}
\label{tab:nucl_charracteristics}
\end{table}
In general, differences among various models within the hyperonic sector become more pronounced at low temperatures and higher densities,
rather than in dilute and hot matter because they arise mainly from the
modeling of the interactions. Furthermore, the model uncertainties
present in the purely nucleonic sector propagate in the hyperonic sector
through the mutual dependence of the baryon octet chemical potentials
imposed by baryon conservation and charge neutrality. To give an
example, we note that a small magnitude of the symmetry energy of
nucleonic matter disfavors hyperons~\cite{Fortin2018}. As mentioned
above, the requirement that the maximum mass of a cold hypernuclear
star be larger than the observational limit $2 M_\odot$, favors
nucleonic models with hard EoS and hyperonic interactions that become
sufficiently repulsive at high densities. These features have been
discussed extensively in the context of the ``hyperon puzzle", for reviews
see~\cite{Chatterjee2016,Sedrakian2023}.

Let us start our discussion with the nucleonic parametrizations that
have been used as a basis for extensions to the hypernuclear sector.
We list their nuclear characteristics
in Table~\ref{tab:nucl_charracteristics}. They can be divided into several classes. \\
{\it 1. CDFs with density-dependent couplings.}
Refs.~\cite{Marques2017,Fortin2018,Raduta2020,Raduta2022} developed
models based on the DD2~\cite{Typel2010} parameterization and their
extension to the hypernuclear sector including the full baryon octet;
some of the models include also
$\Delta$-resonances~\cite{Raduta2020,Raduta2022}. The hypernuclear
models of Refs.~\cite{Banik2014} based on the same parametrization
include $\Lambda$ hyperons only. The nucleonic DD2
model~\cite{Typel2010} predicts a moderate value of the slope of the
symmetry energy $\Lsym=55$~MeV and standard values of the remaining
characteristics of nuclear matter. Among our models, the DDLS(50) model
thus has a similar value $\Lsym=50$ MeV. However, there are stronger variations
in the values of $\Qsat$, specifically $\Qsat=479.22$~MeV for the
DDME2 model~\cite{Li2019}, $\Qsat=169.15$~MeV for the DD2 model, and
$\Qsat=400$~MeV in the present work. All models guarantee that
hyperonic stars have masses larger than the two-solar limit; present
models allow to vary $\Lsym$ parameter and explore its influence on
various astrophysical scenarios.

{\it 2. Quark-meson coupling model.} Ref.~\cite{Stone2021} generated
tables for these types of models that include a hyperonic component;
these are labeled in the {\sc CompOSE} database as SDGTT(QMC-A). As
can be seen from Table~\ref{tab:nucl_charracteristics}, the slope of
the symmetry energy is close to the DDLS(50) model. However, the
symmetry energy at saturation is smaller compared DDLS(50) model and
other models discussed above,
which implies that its value at the crossing density of $n_B = 0.110$
fm$^{-3}$ is likewise smaller than for the rest of the model collection.
Another notable difference is that the compressibility of
nuclear matter is by about 20\% larger than in the models used here.

{\it 3. Chiral mean-field (CMF) model.} This model [labeled in {\sc CompOSE}
database as DNS(CMF)] is based on a non-linear realization of the
sigma model which includes pseudo-scalar mesons as the angular
parameters for the chiral
transformation~\cite{Schurhoff2010,Dexheimer2015}. This model has
large compressibility (which is comparable to the QMC model) and a
value of $\Lsym$ which is by about $25\%$ larger than the value for
$DDLS(70)$ model. The model is characterized by a low-value symmetry
energy at saturation and its steep increase at higher densities. It
appears that the density-dependence of the symmetry energy of the CMF model differs considerably from other models except for the QMC
model.

{\it 4. Non-linear mean-field models.} These models are based on the CDFs with nonlinear meson self-interactions. The FSU2H parametrization was specifically
designed for an extension to the hypernuclear sector~\cite{Tolos2017},
see also its updated version in Ref.~\cite{Kochankovski2022} labeled
in the {\sc CompOSE} database as KRT(FSU2H*). The alternative
SFHo~\cite{Steiner2013} parametrization was extended to hypernuclear
sector in Ref.~\cite{Fortin2018} and is labeled in the database as
OMHN(DD2Y). For both models, the nuclear parameters are close to their central values; in particular, the $\Lsym$ value is close to the DDLS(50) model used in this work.

In Fig.~\ref{fig:hyperonfraction} we display the total hyperon
fraction $Y_H = \sum_{i} Y_i$, where the index $i$ sums over the
hyperons, as a function of baryon number density for the same
temperatures $T=10$ and 80 MeV and electron fractions $Y_e=0.1$ and
$0.4$ as in Figs.~\ref{fig:abundances-T_const_Ye01} and
\ref{fig:abundances-T_const_Ye04}.  As already seen above, there is
only a slight difference in the hyperon content of the three models
based on DDLS(30,50,70) parametrization. The value of $Y_H$ is
slightly larger for a larger value of $\Lsym$.  We anticipate that the
variations of the slope of the symmetry energy at saturation density
has little influence on the hyperonic content of the models which is
determined by the physics at higher densities, where the contribution
of the $\vecrho$-meson responsible for variations in $\Lsym$ is
exponentially suppressed, see Eq.~\eqref{eq:h_function_rho}.  Note
also that the magnitude of $Y_H$ at low temperatures is dominated by
the $\Lambda$ hyperons independent of the value of the electron
fraction.

At low temperatures, our models and the model BHB($\Lambda\phi$) of
Ref.~\cite{Banik2014} predict the largest hyperon fractions which are
nearly identical.  Slightly smaller hyperon fractions are predicted by
the models OMHN(DD2Y) of Ref.~\cite{Marques2017} based on DD2
parametrization, model KRT(FSU2H*) of Ref.~\cite{Kochankovski2022}
based on FSU2H* parametrization, and model
FOP(SFHoY) \cite{Fortin2018} based on SFHo parametrization (maximally
about 15\% reduction). This difference does not change with
temperature, as it reflects the differences in the hyperonic
parametrizations. At high temperatures, the BHB($\Lambda\phi$) model
result deviates more strongly from the other models as it contains
only $\Lambda$ hyperons. This underlines the importance of including
hyperon species other than the $\Lambda$ at high temperatures.  The
density range where $Y_H$ is non-negligible changes with temperature:
In the low-temperature regime, it abruptly increases above the
threshold density of hyperons, whereas in the high-temperature regime
it extends to much lower densities. While all models exhibit these
features, the density dependence, and the magnitude of $Y_H$ predicted
by the DNS(CMF)~\cite{Schurhoff2010,Dexheimer2015} model and
SDGTT(QMC-A)~\cite{Stone2021} model are markedly different at low
temperatures and there are still some qualitative differences at high
temperatures. Our brief review in this section highlights 
the necessity of a more detailed comparison of the predictions of the
modern CDF models in the hyperonic sector. Uncovering the origin of
the discrepancies among the CDFs remain an interesting task for the
future.

\section{Conclusions}
\label{sec:Conclusions}

In this work, we have computed the EoS and composition of finite
temperature nucleonic and hypernuclear matter on three-dimensional
grids of temperatures, densities, and electron fractions that are
required by input tables for simulations of CCSN and BNS mergers. The 
homogeneous matter was computed down to $0.5 n_{\text{sat}}$  
and the extension to lower densities was done by matching  to the
low-density HS(DD2) model. The
high-density EoS is based on the DDLS family of parametrizations~\cite{Li2023} which
allow (in general) variations of the slope of the symmetry energy
$L_{\rm sym}$ and skewness $Q_{\rm sat}$; we have chosen to work with
$Q_{\rm sat}=400$~MeV which guarantees that the values of
maximum masses of hypernuclear stars are above the two solar mass
lower bound and
changed the value of $L_{\rm sym} = 30,$ 50 and 70 MeV. The range of
$L_{\rm sym} $ has been chosen to allow for the current uncertainties
in this quantity associated with the measurements of neutron skin 
experiments as well as radii of neutron stars. Thus, our collection
of EoS, among other things, allows one to study the effects of varying the symmetry energy in dynamical transients such as supernova explosions and BNS mergers.  The six EoS and composition tables that were generated are available on the {\sc CompOSE}  database.

\section*{Acknowledgments}
S.~T. and A. S. acknowledge the support of the Polish National Science
Centre (NCN) grant 2020/\-37/B/ST9/01937.
S.~T. is supported by the IMPRS for Quantum Dynamics and Control at
Max Planck Institute for the Physics of Complex Systems, Dresden,
Germany.  A.~S. acknowledges Grant No. SE 1836/5-2 from Deutsche
For\-schungs\-gemeinschaft (DFG) and useful communications with J.~J.~Li. M.~O. acknowledges financial support
from the Agence Nationale de la recherche (ANR) under contract number
ANR-22-CE31-0001-01.

\bibliographystyle{JHEP}
\providecommand{\href}[2]{#2}\begingroup\raggedright\endgroup

\end{document}